\documentclass[a4paper,11pt]{article}
\usepackage{jheppub} 
\usepackage[T1]{fontenc} 
\RequirePackage[displaymath]{lineno} 
\usepackage{epsfig}
\usepackage{graphicx}
\usepackage{dcolumn}
\usepackage{bm}
\usepackage{ltablex,booktabs}
\usepackage{overpic}
\usepackage{subfigure}
\usepackage{amsmath}
\usepackage{amsmath}
\usepackage{float}
\usepackage{color}
\usepackage{amsmath}
\usepackage{mathcomp}
\usepackage{mathrsfs}
\usepackage{multirow}
\usepackage{rotating}
\usepackage{amssymb}
\usepackage{gensymb}
\usepackage{amsmath}
\usepackage{tabularx}
\usepackage{epstopdf}
\usepackage{tikz-feynman}
\usepackage{makecell}
\usepackage{relsize}
\usepackage{xcolor}
\usepackage{pdflscape}
\def\babar{\mbox{\slshape B\kern-0.1em{\smaller A}\kern-0.1em B\kern-0.1em{\smaller A\kern-0.2em R}}}

\def\BF{\ensuremath{(0.73\pm0.03_{\rm stat.}\pm0.01_{\rm syst.})\times10^{-3}}}

\title{Amplitude analysis and branching fraction measurement of the decay \boldmath $D^0 \to K^+K^-\pi^0\pi^0$}
\collaborationImg{\includegraphics[width=0.15\textwidth, angle=90]{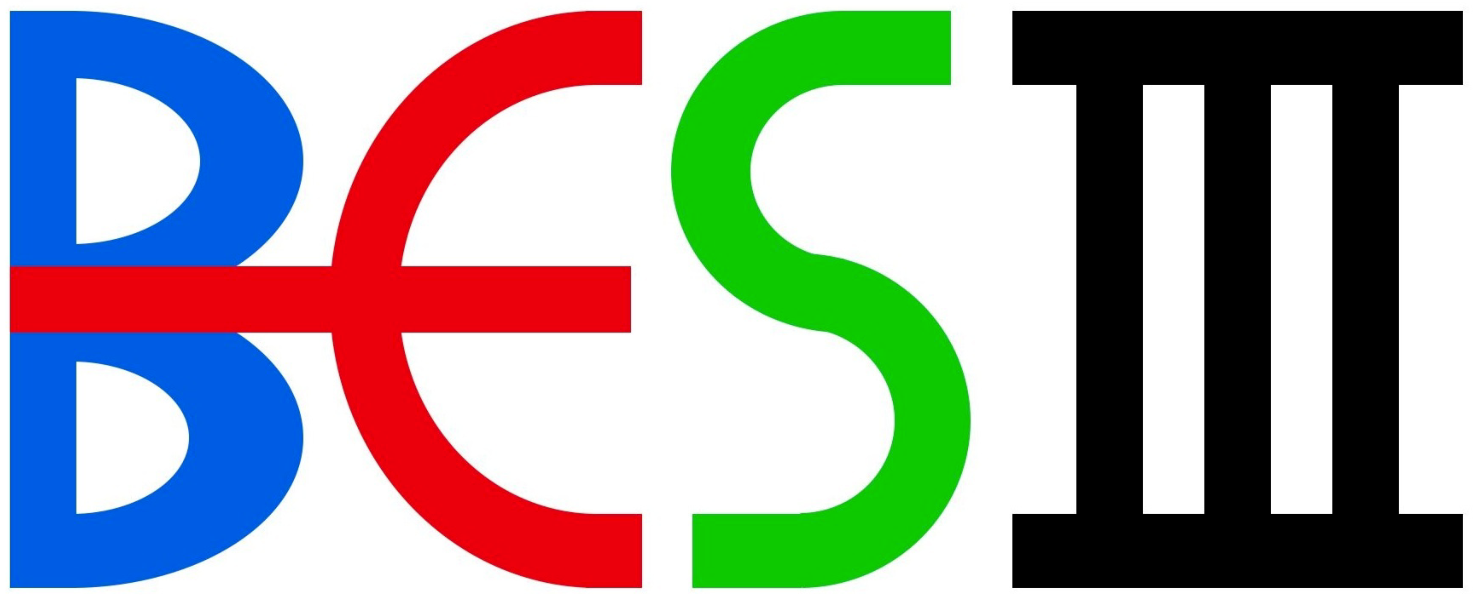}}
\collaboration{BESIII Collaboration}
\date{\today}
\abstract{
An amplitude analysis of the singly Cabibbo-suppressed decay $D^0 \to K^+ K^- \pi^0 \pi^0$ is performed, for the first time, to determine the relative magnitudes and phases of different intermediate processes. The analysis uses $e^+e^-$ collision data collected with the BESIII detector at the center-of-mass energy 3.773~GeV corresponding to an integrated luminosity of 20.3 $\rm fb^{-1}$. 
The absolute branching fraction of $D^0 \to K^+ K^- \pi^0 \pi^0$ is measured to be \BF. The dominant intermediate process is $D^0 \to K^{*}(892)^+K^{*}(892)^-$, with a branching fraction of $(2.79 \pm 0.13_{\rm{stat.}} \pm 0.11_{\rm{syst.}}) \times 10^{-3}$. 
Amplitude analysis reveals that the $D^0 \to K^{*}(892)^+K^{*}(892)^-$ decay is S-wave dominant. The longitudinal polarization fraction of $D^0 \to K^{*}(892)^+ K^{*}(892)^-$ is measured to be $0.468\pm0.046_{\rm{stat.}}\pm0.011_{\rm{syst.}}$.}
\keywords{Amplitude Analysis, Charm Physics, $e^+e^-$ Collider Experiment, Branching Fraction}
\arxivnumber{}
\begin{document}
\newcommand{\BESIIIorcid}[1]{\href{https://orcid.org/#1}{\hspace*{0.1em}\raisebox{-0.45ex}{\includegraphics[width=1em]{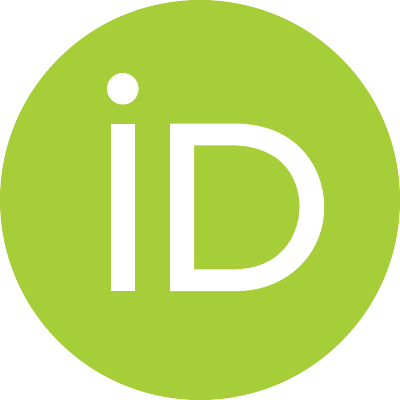}}}}
\maketitle
\flushbottom

\section{Introduction}
Theoretical studies of hadronic decays of charm mesons are challenging because the charm quark mass is neither heavy enough for a reliable heavy quark mass expansion, nor light enough for the application of chiral perturbation theory~\cite{Cheng}. To address these challenges, non-perturbative methods are employed~\cite{Bajc_1997}, which rely on precise experimental inputs to constrain model parameters, test theoretical predictions and guide the refinement of theoretical frameworks. This close interplay between theory and experiment drives the progress in understanding $D$-meson decays, particularly in studies of \emph{CP} violation in $D$ decays~\cite{TDA_QCD}.

\begin{figure}[t]
\centering
    \subfigure[]{\includegraphics[width=0.48\textwidth]{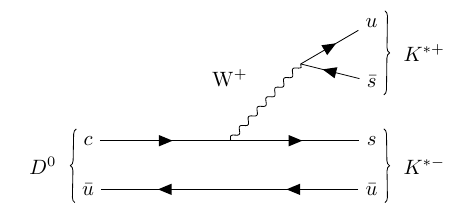}}
    \subfigure[]{\includegraphics[width=0.48\textwidth]{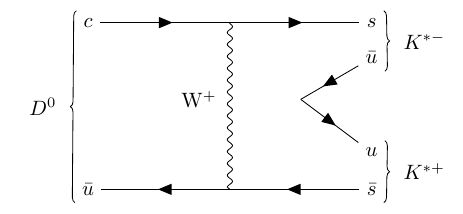}}
    \caption{Topological diagrams contributing to the decay $D^0 \rightarrow  K^{*}(892)^+K^{*}(892)^-$ with (a) color-allowed external W-emission tree diagram and (b) color-suppressed internal $W$ emission and $W$-exchange diagram.}
\label{topology1}
\end{figure}

The lightest charmed mesons, $D^{0}$ and $D^{\pm}$, decay exclusively through the weak interaction. Their decay amplitudes are primarily governed by two-body processes such as $D \rightarrow VP$, $D \rightarrow PP$, $D \rightarrow AP$, and $D \rightarrow VV$, where $V$, $A$ and $P$ represent vector, axial vector and pseudoscalar mesons, respectively. The decay $D^0 \to K^+K^-\pi^0\pi^0$ is a singly Cabibbo-suppressed (SCS) decay, with a major contribution expected from the $D \rightarrow VV$ decay process $D^0 \to K^{*}(892)^+ K^{*}(892)^-$, as illustrated in Fig.~\ref{topology1}. This decay can proceed via the color-allowed external W-emission topology and via color-suppressed mechanisms, including $W$-exchange between the quark and anti-quark inside the $D^0$.
To better understand these decay mechanisms, various phenomenological approaches have been developed, such as the non-relativistic constituent quark model~(NRCQM) with or without final state interactions (FSIs)~\cite{Cao:2023csx}, the flavor SU(3) symmetry model~\cite{Kamal:1990ky}, and the factorization approach~\cite{Cheng:2010rv}, which describe several $D \rightarrow VV$ decay channels well. The predicted values of the absolute $\mathcal{B} (D^0\to  K^{*}(892)^+K^{*}(892)^-)$ are listed in Table~\ref{topology1}. 

\begin{table}[htbp]
  \centering
  \begin{tabular}{|lc|}
    \hline
    Model  &$\mathcal{B} (D^0\to  K^{*}(892)^+K^{*}(892)^-)$~($\times 10^{-3}$)\\
    \hline	
    NRCQM with FSIs~\cite{Cao:2023csx}
    &9.92$\pm$0.34 \\
    NRCQM without FSIs~\cite{Cao:2023csx}
    &5.86 \\
    Flavor SU(3) symmetry model~\cite{Kamal:1990ky} 
    &2.4\\
    Factorization approach~\cite{Cheng:2010rv}
    &7.3\\
    \hline
  \end{tabular}
  \caption{Theoretical predictions for $\mathcal{B}(D^0 \to K^{*}(892)^+K^{*}(892)^-)$.} 
  \label{tab:compare}
\end{table} 

In the $D \rightarrow VV$ decay, the vector mesons can be produced in three polarization states. Consequently, the decay $D^0 \rightarrow VV$ generates one longitudinal amplitude ($H_0$) and two transverse amplitudes ($H_\pm$). The longitudinal polarization fractions, $F_L = |H_0|^2/|H_-+H_0+H_+|^2$, in heavy flavor meson decays to $VV$ final states have long been of significant interest, as this quantity uniquely provides sensitivity to potential new physics effects and insights into the FSI dynamics~\cite{Dunietz:1990cj,Valencia:1988it}. Both non-relativistic constituent quark models, with and without FSIs, predict transverse-polarization dominance in the $D^0 \to K^{*}(892)^+ K^{*}(892)^-$ process, with longitudinal polarization fractions of $0.32^{+0.02}_{-0.01}$ and 0.31, respectively~\cite{Cao:2023csx}.The measurement of the decay $D^0 \to K^{*}(892)^+ K^{*}(892)^-$ will provide the first experimental measurement of its polarization fractions, thus testing the predictions of these models. Furthermore, the $D^0 \to K^+K^-\pi^0\pi^0$ decay also encompasses numerous intermediate $D \to AP$ processes, which provide unique information to study axial-vector mesons in SCS decay, such as $K_1(1270)$, $K_1(1400)$, $f_1(1420)$, etc.~\cite{Cheng:2013cwa,Cheng:2011pb}. 

This paper presents an amplitude analysis and branching fraction (BF) measurement of the decay $D^0 \to K^+K^-\pi^0\pi^0$, utilizing $e^+e^-$ collision data collected by the BESIII detector at the center-of-energy 3.773~GeV corresponding to an integrated luminosity of 20.3 $\rm fb^{-1}$~\cite{lum1,lum2}. Charged-conjugate modes are always implied throughout this paper.

\section{Detector and data}
\label{sec:detector_dataset}
The BESIII  detector~\cite{Ablikim:2009aa} records symmetric $e^+e^-$ collisions provided by the BEPCII storage ring~\cite{Yu:IPAC2016-TUYA01} in the center-of-mass energy range from 1.84 to 4.95~GeV, with a peak luminosity of $1.1 \times 10^{33}\;\text{cm}^{-2}\text{s}^{-1}$ achieved at $\sqrt{s} = 3.773\;\text{GeV}$. BESIII has collected large data samples in this energy region~\cite{Ablikim:2019hff, EcmsMea, EventFilter}. The cylindrical core of the BESIII detector covers 93\% of the full solid angle and consists of a helium-based  multilayer drift chamber~(MDC), a plastic scintillator time-of-flight system~(TOF), and a CsI(Tl) electromagnetic calorimeter~(EMC), which are all enclosed in a superconducting solenoidal magnet providing a 1.0~T magnetic field. The solenoid is supported by an octagonal flux-return yoke with resistive plate counter muon identification modules interleaved with steel. The charged-particle momentum resolution at 1 GeV$/c$ is $0.5\%$, and the ionization energy loss (${\rm d}E/{\rm d}x$) resolution in the MDC is $6\%$ for electrons from Bhabha scattering. The EMC measures photon energies with a resolution of $2.5\%$ ($5\%$) at $1$~GeV in the barrel (end-cap) region. The time resolution in the TOF barrel region is 68~ps, while that in the end-cap region is 110~ps.  The end-cap TOF system was upgraded in 2015 using multigap resistive plate chamber technology, providing a time resolution of 60~ps, which benefits 86\% of the data used in this analysis~\cite{etof1,etof2,etof3}.

The data sample, with a total integrated luminosity of $20.3 \, \mathrm{fb}^{-1}$, collected at a center-of-mass energy of $\sqrt{s} = 3.773\, \mathrm{GeV}$, is used in this analysis. Here, the $\psi(3770)$ predominantly decays to $D^+D^-$ or $\bar{D}^0 D^0$ pairs without additional hadronic background. This provides an ideal environment for studying $D$ meson decays with the double-tag (DT) technique~\cite{DTmethod}. For a DT candidate, both the $D^0$ and $\bar D^0$ mesons are reconstructed, with the $D^0$ meson decaying to the signal mode $D^0 \to K^+K^-\pi^0\pi^0$ and the $\bar D^0$ meson decaying to one of the reconstructed single-tag (ST ) modes: $\bar D^0 \to K^+\pi^-$, $\bar D^0 \to K^+\pi^-\pi^0$ and $\bar D^0 \to K^+\pi^-\pi^-\pi^+$. 

Monte Carlo (MC) simulated data samples produced with a {\sc geant4}-based~\cite{geant4} software package, which includes the geometric description of the BESIII detector its response, are used to determine detection efficiencies and estimate backgrounds. The simulation models the beam-energy spread and initial-state radiation (ISR) in the $e^+e^-$ annihilations with the generator {\sc kkmc}~\cite{KKMC1,KKMC2}. The inclusive MC sample includes the production of $D\bar{D}$ pairs (with quantum-coherence effects for the neutral $D^0 \bar D^0$ channels), the non-$D\bar{D}$ decays of the $\psi(3770)$, the ISR production of the $J/\psi$ and $\psi(3686)$ states, and the continuum processes incorporated in {\sc kkmc}~\cite{KKMC1,KKMC2}. All particle decays are modeled with {\sc evtgen}~\cite{EVTGEN1,EVTGEN2} using BFs either taken from the Particle Data Group (PDG)~\cite{PDG}, when available, or otherwise estimated with {\sc lundcharm}~\cite{LUNDCHARM1,LUNDCHARM2}. Final-state radiation from charged final-state particles is incorporated using {\sc photos}~\cite{PHOTOS}. A phase-space (PHSP) MC sample is generated with events uniformly distributed over the $D^0 \to K^+K^-\pi^0\pi^0$ phase space, which is used to determine the efficiency function described in Sec.~\ref{sec:fitmethod} and calculate the normalization integral used in the determination of the amplitude-model parameters in the fit to data. According to the results of the amplitude analysis, a signal MC sample is generated. The signal MC is used to check the fit performance, calculate the goodness of fit and estimate the detection efficiency.

\section{Event selection}
\label{ST-selection}
The $D^0$ candidates are constructed from individual $\pi^\pm$, $\pi^0$ and $K^\pm$ candidates with selection criteria that are common for both the amplitude analysis and the BF measurement.  
 
Charged tracks detected in the MDC are required to be within a polar angle ($\theta$) range of $|\rm{cos\theta}|<0.93$, where $\theta$ is defined with respect to the $z$-axis, which is the symmetry axis of the MDC. For charged tracks, the distance of closest approach to the interaction point (IP) must be less than 10\,cm along the $z$-axis,  and less than 1\,cm in the transverse plane.

Particle identification~(PID) for charged tracks combines measurements of the ionization energy loss d$E$/d$x$ and the flight time in the TOF to form likelihoods $\mathcal{L}(h)~(h=K,\pi)$ for each hadron $h$ hypothesis.
Charged kaons and pions are identified by comparing the likelihoods, $\mathcal{L}(K)>\mathcal{L}(\pi)$ and $\mathcal{L}(\pi)>\mathcal{L}(K)$, respectively.

Photon candidates are identified using isolated showers in the EMC.  The deposited energy of each shower must be more than 25~MeV in the barrel region ($|\!\cos \theta|< 0.80$) and more than 50~MeV in the end-cap region ($0.86 <|\!\cos \theta|< 0.92$).  To exclude showers that originate from charged tracks, the angle between the EMC shower and the position of the closest charged track at the EMC must be greater than 10 degrees as measured from the IP. To suppress electronic noise and showers unrelated to the event, the difference between the EMC time and the event start time is required to be within [0, 700]\,ns.

The $\pi^0$ candidates are formed from the photon pairs with invariant masses in a range of $[0.115, 0.150]$~GeV/$c^2$, which is about three times the mass resolution. Moreover, to achieve an adequate resolution, at least one of the two photons must be detected in the barrel EMC. A kinematic fit that constrains the $\gamma\gamma$ invariant mass to the known $\pi^{0}$ mass~\cite{PDG} is performed to improve the mass resolution. The $\chi^2$ of the kinematic fit is required to be less than 50.

To distinguish the $D^0 \bar D^0$ mesons from the backgrounds, the beam-constrained mass~($M_{\rm{BC}}$) and the energy difference~($\Delta E$) are used to identify the signal $D^0 \bar D^0$ pair,
\begin{equation}
\begin{aligned}
	&M_{\rm{BC}} = \sqrt{E_{\rm{beam}}^2 /c^4-|\vec{p}_{D}|^2 /c^2}, \\
	&\Delta E = E_{D} - E_{\rm{beam}},
\end{aligned}
\label{MBC_DeltaE}
\end{equation}
where $\vec{p}_{D}$ and $E_{D}$ are the total reconstructed momentum and energy of the $D$ candidate, and $E_{\rm{beam}}$ is the beam energy. The $D$ signal manifests itself as a peak around the known $D$ mass~\cite{PDG} in the $M_{\rm{BC}}$ distribution and as a peak around zero in the $\Delta E$ distribution. For each tag mode, if there are multiple combinations, the one giving the minimum $|\Delta E_{\text{tag}}|$ is retained for further analysis. The signal $D^0$ candidates are reconstructed from the particles that are not used for the tagged $\bar D^0$ reconstruction, with $\pi^0$ being reconstructed from $\gamma \gamma$. They are identified using the energy difference and the beam-constrained mass of the signal~(sig) side, $|\Delta E_{\text{sig}}|$ and $M^{\text{sig}}_{\mathrm{BC}}$. If there are multiple combinations, the one giving the minimum $|\Delta E_{\text{sig}}|$ is retained for further analysis.

In order to enhance the selection of $D$ mesons and effectively suppress background events, mode-dependent requirements on the energy difference $\Delta E$ are applied to both the tag modes. The corresponding $\Delta{E}$ regions of the signal and tag modes are provided in Table~\ref{tab:deltaEcut}.

\begin{table}[hbtp]
  \begin{center}
     \centering \begin{tabular}{|lc|}
      \hline
      Decay mode &$\Delta{E}~(\rm {GeV})$\\
      \hline
      $D^0\to K^+K^- \pi^0\pi^0$ &(-0.020, 0.020)\\
      \hline
      $\bar {D}^0\to K^+\pi^-$  &(-0.025, 0.025)\\
      $\bar {D}^0\to K^+\pi^-\pi^0$  &(-0.055, 0.040)\\
      $\bar {D}^0\to K^+\pi^-\pi^-\pi^+$  &(-0.025, 0.025)\\
      \hline
    \end{tabular}
    \caption{Requirements of $\Delta{E}$ for the signal and the different $\bar D^0$ tag modes.}
    \label{tab:deltaEcut}
  \end{center}
\end{table}

Subsequently, the sources of background in the selection of  $D^0 \to K^+K^- \pi^0 \pi^0$ candidates are investigated by analyzing the inclusive MC sample. A $K^0_S$ mass veto, $M_{\pi^0 \pi^0}$ $\notin[0.460, 0.525]~{\rm GeV}/c^2$, which is about $\pm3.5$ times of the $K^0_S$ resolution, is applied on the signal $D^0$ to remove the dominant background from $D^0\to K^0_S K^+K^-$ decays. To suppress non-$D^0\bar{D}^0$ backgrounds which mainly come from the candidate events with correctly reconstructed $D^0(\bar{D}^0)$ and incorrectly reconstructed $\bar{D}^0(D^0)$, the opening angle between these two $D$ mesons ($\theta_{D^0\bar{D}^0}$) is required to be greater than $167^\circ$. With this condition, 98.8\% signal events are kept, and 67.9\% background events are rejected.

\section{Amplitude analysis}
\label{Amplitude-Analysis}
 \subsection{Signal extraction for amplitude analysis}
\label{sec:pwa-select}

To improve the signal purity for the amplitude analysis, events in the signal region are selected based on the criterion $1.861 < M_{\rm BC} < 1.870~{\rm GeV}/c^2$ for both the signal and tag modes. For optimal resolution and to ensure that all events are within the PHSP boundary, a full-constraint kinematic fit is performed, in which the four-momenta of the final-state particles are constrained to the initial four-momenta of the $e^+e^-$ system and the reconstructed masses of the $D^0$ and $\pi^0$ mesons are constrained to their known values~\cite{PDG}. These four-momenta are used for the amplitude analysis. 

\begin{figure}[htbp]
  \centering
 \includegraphics[trim=0 0 0 0,width=0.90\textwidth]{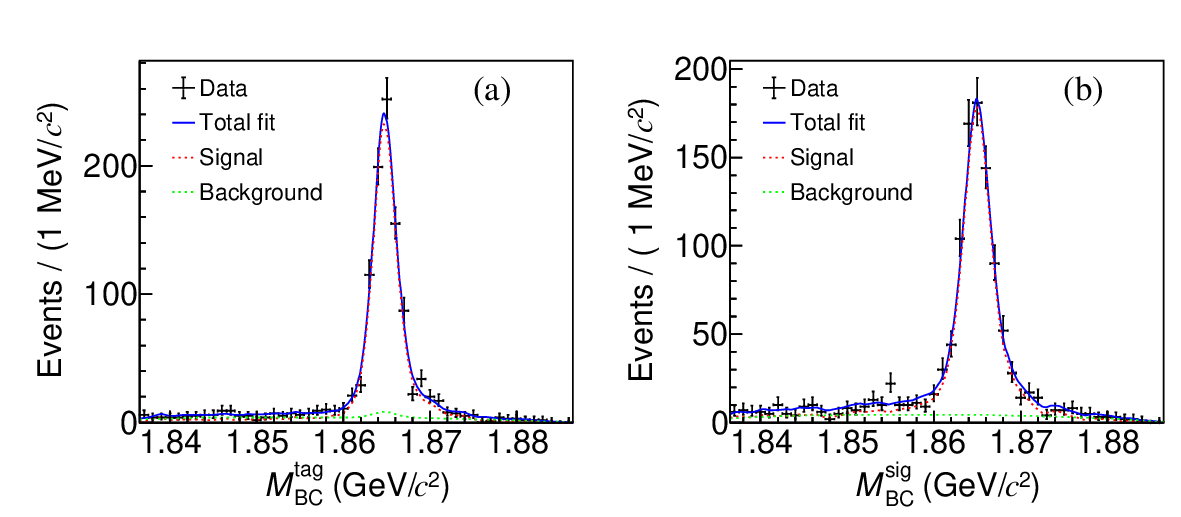}
  \caption{Projections on (a)~$M_{\rm BC}^{\rm tag}$ and (b)~$M_{\rm BC}^{\rm sig}$ from the 2D fit. The points with error bars are data. The solid blue lines are the total fit result, the dotted red lines are the signal, and the dotted green lines are the background. }
  \label{fig:2dfit}
\end{figure}

After applying all of the aforementioned criteria, a total of 791 events are retained in the signal region for the subsequent amplitude analysis. The signal purity ($\omega_{\rm{sig}}$), determined from an unbinned 2D maximum likelihood fit to the $M_{\rm{BC}}^{\rm{sig}}$ versus $M_{\rm{BC}}^{\rm{tag}}$ distribution (see Appendix~\ref{2dfit} for details), is measured to be $(94.2\pm 0.2)\%$. The uncertainty in purity is obtained by propagating the uncertainties of the fitting parameters according to the correlation matrix. The fit results are shown in Fig.~\ref{fig:2dfit}.

\subsection{Fit method}
\label{sec:fitmethod}
An unbinned maximum-likelihood fit is used in the amplitude analysis of $D^0 \to K^+K^-   \pi^0\pi^0$. The likelihood function $\mathcal{L}$ is constructed by incoherently adding the background probability density function (PDF) to the signal PDF. After taking the logarithm, the log-likelihood is written as

\begin{eqnarray}\begin{aligned}
\ln{\mathcal{L}} = \begin{matrix}\sum\limits_{k=1}^{N_{\rm data}} \ln [\omega_{\rm sig}f_{S}(p^{k})+(1-\omega_{\rm sig})f_B(p^k)]\end{matrix}\,,
\label{likelihood3}
\end{aligned}\end{eqnarray}
where $k$ indicates the $k^{\rm th}$ event in the data sample, $N_{\rm data}$ is the number of retained events, $p$ denotes the four-momenta of the final state particles, $f_{S}(f_{B})$ is the signal (background) PDF and $\omega_{\rm sig}$ is the signal purity discussed in Sec.~\ref{sec:pwa-select}.

The signal PDF $f_S$ is given by 
\begin{eqnarray}\begin{aligned}
  f_S(p) = \frac{\epsilon(p)\left|\mathcal{M}(p)\right|^{2}R_{4}(p)}{\int \epsilon(p)\left|\mathcal{M}(p)\right|^{2}R_{4}(p)\,{\rm d}p}\,, \label{signal-PDF}
\end{aligned}\end{eqnarray}
where $\epsilon(p)$ is the detection efficiency and $R_4(p)$ is the PHSP factor of four-body decays. The total amplitude $\mathcal{M}$ is treated within the isobar model, which uses the coherent sum of amplitudes of intermediate processes, given by $\mathcal M(p) = \sum{c_n\mathcal A_n(p)}$, where $c_n = \rho_ne^{i\phi_n}$ and $\mathcal A_n(p)$ are the complex coefficient and the amplitude for the $n^{\rm th}$ intermediate process, respectively. The magnitude $\rho_n$ and phase $\phi_n$ are free parameters in the fit. We use covariant tensors to construct amplitudes, which are written as
\begin{equation}
  \mathcal A_n(p_j) = P_n^1P_n^2S_nX_n^1X_n^2X_n^{D^0},
\end{equation}

where $S_n$ and { $X_n^{1,2}(X_n^{D^0})$} are the spin factor and the Blatt-Weisskopf barriers for the intermediate resonances (the $D^0$ meson), respectively, (Sec.~\ref{Blatte}), $P_{n}$ is the propagator of the intermediate resonance (Sec.~\ref{sec:propagator}), and $S_{n}$ is the spin factor constructed with the covariant tensor formalism~\cite{covariant-tensors} (Sec.~\ref{sec:spinfactor}).

The background PDF is given by
\begin{equation}
  f_B(p)=\frac{\epsilon(p)B_{\epsilon}(p)R_4(p)}{\int{\epsilon(p)B_{\epsilon}(p)R_4(p)}dp},	
  \label{pdf:bkg_new}
\end{equation}
where $B_{\epsilon}(p) = B(p) / \epsilon(p)$  is the efficiency-corrected background function, and $B(p)$ is the observed background distribution. The background function in the data is modeled by the background events in the signal region derived from the inclusive MC sample. The invariant-mass distributions of two and three track combinations outside the signal region show good agreement between the data and MC simulation, thus validating the description from the inclusive MC sample. The distributions of background events from the inclusive MC sample have been examined both inside and outside the signal region. They are generally compatible with each other within statistical uncertainties.

The invariant-mass distributions $(M_{K^+K^-}, M_{K^+\pi^0_1}, M_{K^+\pi^0_2}, M_{K^-\pi^0_1}, M_{K^-\pi^0_2}, M_{\pi_1 ^0 \pi_2 ^0},\\ M_{K^+K^-\pi_1^0}, M_{K^+K^-\pi_2^0}, M_{K^+\pi_1^0\pi_2^0}, M_{K^-\pi_1^0\pi_2^0})$, and helicity angles, and decay plane angles distributions of background events in the inclusive MC sample and signal events in PHSP MC sample are modeled using the XGBoost package~\cite{XG1,XG2}. The two $\pi^0$ candidates are distinguished based on their momenta, with $\pi^0_1$ having a higher momentum than $\pi^0_2$. The first XGBoost model is used to predict the background probability $B(p)R_4(p)$, where the training samples consist of background events from the inclusive MC sample. Subsequently, a separate XGBoost model is employed to predict the efficiency probability $\epsilon(p)R_4(p)$, with the training dataset derived from PHSP MC samples. Dividing the two probabilities results in the value of $B_{\epsilon}(p)$, which is equal to $B(p)/\epsilon(p)$.

By combining Eq.~(\ref{signal-PDF}) with Eq.~(\ref{pdf:bkg_new}) and neglecting the constant term $\epsilon(p)R_4(p)$, the log-likelihood becomes
\begin{equation}
\ln\mathcal{L}=\sum\limits_{k}^{N_{\rm {data}}}\ln\bigg[\omega_{\rm {sig}}\frac{|M(p)|^2}{\int \epsilon(p)|M(p)|^2R_4(p)dp}+(1-\omega_{\rm {sig}})\frac{B_{\epsilon}(p)}{\int{\epsilon(p)B_{\epsilon}(p)R_4(p)}dp}\bigg].
  \label{likelihoodexa}
\end{equation}

The normalization integrals of signal and background are evaluated with signal MC samples,
\begin{eqnarray}\begin{aligned}
  \int \epsilon(p) |\mathcal{M}(p)|^2 R_{4}(p)\,{\rm d}p \propto
\frac{1}{N_{\rm MC}}\sum_{k_{\rm MC}}^{N_{\rm MC}} \frac{ |\mathcal{M}(p^{k_{\rm MC}})|^2 }{\left|\mathcal{M}^{g}(p^{k_{\rm MC}})\right|^{2}}\,, \label{MC-intergral}
\end{aligned}\end{eqnarray}

\begin{eqnarray}\begin{aligned}
  \int \epsilon(p) B_{\epsilon}(p) R_{4}(p)\,{\rm d}p \propto
\frac{1}{N_{\rm MC}}\sum_{k_{\rm MC}}^{N_{\rm MC}} \frac{ B_{\epsilon}(p^{k_{\rm MC}}) }{\left|\mathcal{M}^{g}(p^{k_{\rm MC}})\right|^{2}}\,, \label{bkgintergral}
\end{aligned}\end{eqnarray}
where $k_{\rm MC}$ is the index of the $k^{\rm th}$ event of the signal MC sample, and $N_{\rm MC}$ is the number of the selected signal MC events. The symbol $\mathcal{M}^{g}(p)$ denotes the PDF used to generate the signal MC sample in the MC integration. The computational efficiency of the MC integration is significantly improved by evaluating the normalization integral with signal MC samples. These samples intrinsically take into account the event selection acceptance and the detection resolution. Tracking, PID, as well as $\pi^0$ reconstruction efficiency differences between data and MC simulation are corrected by multiplying the weight of the MC events by a factor $\gamma_{\epsilon}$, which is calculated as
\begin{equation}
  \gamma_{\epsilon}(p_j) = \prod_{n} \frac{\epsilon_{n,\rm data}(p_j)}{\epsilon_{n,\rm MC}(p_j)},
  \label{pwa:gamma}
\end{equation}
where $n$ refers to tracking, PID, $\pi^0$ reconstruction, $\epsilon_{n,\rm data}(p_j)$ and $\epsilon_{n,\rm MC}(p_j)$ are their efficiencies as a function of the momenta of the daughter particles for data and MC simulation, respectively. Then the MC integration is determined by 
\begin{eqnarray}\begin{aligned}
    &\int \epsilon(p) |\mathcal{M}(p)|^2 R_{4}\,{\rm d}p \propto
&\frac{1}{N_{\rm MC}} \sum_{k_{\rm MC}}^{N_{\rm MC}} \frac{ |\mathcal{M}(p^{k_{\rm MC}})|^2 \gamma_{\epsilon}(p^{k_{\rm MC}})}{\left|\mathcal{M}^{g}(p^{k_{\rm MC}})\right|^{2}}\,,
\label{MC-intergral-corrected}
\end{aligned}\end{eqnarray}

\begin{eqnarray}\begin{aligned}
    &\int \epsilon(p) B_{\epsilon}(p) R_{4}\,{\rm d}p \propto
&\frac{1}{N_{\rm MC}} \sum_{k_{\rm MC}}^{N_{\rm MC}} \frac{ B_{\epsilon}(p^{k_{\rm MC}}) \gamma_{\epsilon}(p^{k_{\rm MC}})}{\left|\mathcal{M}^{g}(p^{k_{\rm MC}})\right|^{2}}\,.
\label{bkg-intergral-corrected}
\end{aligned}\end{eqnarray}

\subsubsection{Blatt-Weisskopf barrier factors}\label{Blatte}
The Blatt-Weisskopf barrier factors $X_L(q)$~\cite{Blatte} are the barrier functions for a two-body decay process $a \to bc$. These functions depend on the angular momentum $L$ and the momenta $q$ of the final-state particle $b$ or $c$ in the rest system of $a$. They are taken as
\begin{eqnarray}
\begin{aligned}
 X_{L=0}(q)&=1,\\
 X_{L=1}(q)&=\sqrt{\frac{z_0^2+1}{z^2+1}},\\
 X_{L=2}(q)&=\sqrt{\frac{z_0^4+3z_0^2+9}{z^4+3z^2+9}}\,,
\end{aligned}
\end{eqnarray}
where $z=qR_r$, $z_0=q_0R_r$, and the momentum $q$ is given by
\begin{eqnarray}
\begin{aligned}
q = \sqrt{\frac{(s_a+s_b-s_c)^2}{4s_a}-s_b}\,. \label{q2}
\end{aligned}
\end{eqnarray}
Here $s_a, s_b, \text{and}~s_c$ are the invariant-masses squared of particles $a, b, \text{and}~c$, respectively. The value of $q_0$ is that of $q$ when $s_a = m_a^2$, where $m_a$ is the mass of particle $a$. The effective radium of the barrier, $R_r$, is fixed to 3.0~$({\rm GeV}/c)^{-1}$ for the intermediate resonances and 5.0~$({\rm GeV}/c)^{-1}$ for the $D^0$ meson. 
\subsubsection{Propagator}\label{sec:propagator}
The intermediate resonances $K^{*}(892)^{\pm}$, $f_1(1420)$, $K_1(1400)$, $\eta(1405)$ and $\eta(1475)$ are
parameterized with a relativistic Breit-Wigner function,
\begin{eqnarray}\begin{aligned}
    P(m) = \frac{1}{m_{0}^{2} - m^2 - im_{0}\Gamma(m)/c^2}\,,\; 
    \Gamma(m) = \Gamma_{0}\left(\frac{q}{q_{0}}\right)^{2L+1}\left(\frac{m_{0}}{m}\right)X^{2}_{L}(q)\,, 
  \label{RBW}
\end{aligned}\end{eqnarray}
where $m$ is the invariant mass of the decay products, $m_0$ and $\Gamma_0$ are the mass and width of the intermediate resonance that are fixed to their known values~\cite{PDG}. The energy-dependent width is denoted by $\Gamma(m)$. The $X_L(q)$ is the Blatt-Weisskopf barrier factor, defined in Sec.~\ref{Blatte}.

The $K\pi$ $S$-wave is modeled by the LASS parameterization~\cite{KPsnew1}, which is described by a $K_0^{*}(1430)$ Breit-Wigner together with an effective range non-resonant component with a phase shift. It is given by	
\begin{equation}
A(m)=F\sin\delta_Fe^{i\delta_F}+R\sin\delta_Re^{i\delta_R}e^{i2\delta_F},
\end{equation}
with
\begin{equation}
\begin{aligned}
&\delta_F=\phi_F+\cot^{-1}\left[\frac{1}{aq}+\frac{rq}{2}\right], \\
&\delta_R=\phi_R+\tan^{-1}\left[\frac{M_{K^*_0(1430)}\Gamma(m_{K\pi})/c^2}{M_{K^*_0(1430)}^2-m^2_{K\pi}}\right].
\end{aligned}
\end{equation}
The parameters $F$, $\phi_F$~($R$ and $\phi_R$) are the amplitudes and phases of the non-resonant~(resonant) component, respectively. The parameters $a$ and $r$ are the scattering length and effective interaction length, respectively. 
The parameters $M_{K^*_0(1430)}$ and $m_{K\pi}$ are the $K^*_0(1430)$ mass and the invariant mass of  the $K\pi$ system, respectively.
The parameters $M_{K^*_0(1430)}$, $\Gamma$, $F$, $\phi_F$, $R$, $\phi_R$, $a$, and $r$ are fixed to the values obtained from the amplitude analysis of $D^0 \to K_S^0 \pi^+ \pi^-$ performed by the \babar\ and Belle experiments~\cite{KPsnew2}. The values of these parameters are given in Table~\ref{tab:babar}. Amplitude involving the $K\pi$ $S$-wave have a negligible impact and are therefore excluded from the nominal fit.
\begin{table}[htbp]
\setlength{\abovecaptionskip}{0.cm}
\setlength{\belowcaptionskip}{-0.2cm}
  \begin{center}
    \begin{tabular}{|lc|}
			\hline
      Parameter &Value\\
      \hline
      $M_{K^*_0(1430)}$(GeV/$c^2$)        &1.441 $\pm$ 0.002\\
      $\Gamma$(GeV)   &0.193 $\pm$ 0.004\\
      $F$                 &0.96 $\pm$ 0.07\\
      $\phi_F$~($^\circ$)            &0.1 $\pm$ 0.3\\
      $R$                 &1(fixed)\\
      $\phi_R$~($^\circ$)            &$-$109.7 $\pm$ 2.6\\
      $a$ (GeV/$c$)$^{-1}$                &0.113 $\pm$ 0.006\\
      $r$ (GeV/$c$)$^{-1}$                  &$-$33.8 $\pm$ 1.8\\
      \hline
    \end{tabular}
  \end{center}
  \caption{$K\pi$ $S$-wave parameters obtained from a combined amplitude analysis of $D^0\to K_S^0\pi^+\pi^-$ by the \babar~and Belle experiments~\cite{KPsnew2}. The uncertainties are the combined statistical and systematic uncertainties.}
  \label{tab:babar}
\end{table}
\subsubsection{Spin factors}\label{sec:spinfactor}
Due to the limited size of the PHSP, the analysis is restricted to intermediate resonances with spins $J = 0, 1$, and $2$. In the decay process $a \to bc$, exclusive consideration is given to systems with orbital angular momenta $L = 0, 1$, and $2$, as these configurations dominate the observed decay modes. The momenta of the particles $a$, $b$, and $c$ in the $a \to bc$ process are denoted by $p_a$, $p_b$, and $p_c$, respectively. The spin-projection operators~\cite{covariant-tensors} are defined as
\begin{eqnarray}
\begin{aligned}
  &P^{(0)}(a) = 1\,,&(S\rm{-wave})\\
  &P^{(1)}_{\mu\mu^{\prime}}(a) = -g_{\mu\mu^{\prime}}+\frac{p_{a,\mu}p_{a,\mu^{\prime}}}{p_{a}^{2}}\,,&(P\rm{-wave})\\
  &P^{(2)}_{\mu\nu\mu^{\prime}\nu^{\prime}}(a) = \frac{1}{2}(P^{(1)}_{\mu\mu^{\prime}}(a)P^{(1)}_{\nu\nu^{\prime}}(a)+P^{(1)}_{\mu\nu^{\prime}}(a)P^{(1)}_{\nu\mu^{\prime}}(a)) -\frac{1}{3}P^{(1)}_{\mu\nu}(a)P^{(1)}_{\mu^{\prime}\nu^{\prime}}(a)\,.&(D\rm{-wave})
 \label{spin-projection-operators}
\end{aligned}
\end{eqnarray}
The pure orbital angular-momentum covariant tensors are given by 
\begin{eqnarray}
\begin{aligned}
    \tilde{t}^{(0)}_{\mu}(a) &= 1\,,&(S\rm{-wave})\\
    \tilde{t}^{(1)}_{\mu}(a) &= -P^{(1)}_{\mu\mu^{\prime}}(a)r^{\mu^{\prime}}_{a}\,,&(P\rm{-wave})\\
    \tilde{t}^{(2)}_{\mu\nu}(a) &= P^{(2)}_{\mu\nu\mu^{\prime}\nu^{\prime}}(a)r^{\mu^{\prime}}_{a}r^{\nu^{\prime}}_{a}\,,&(D\rm{-wave})
\label{covariant-tensors}
\end{aligned}
\end{eqnarray}
where $r_a = p_b-p_c$. The spin factors $S(p)$ used in this work are constructed from the spin projection operators and pure orbital angular-momentum covariant tensors and are listed in Table~\ref{table:spin_factors}. The tensor describing the $D^0$ decays with orbital angular-momentum quantum number $l$ is denoted by $\tilde{T}^{(l)\mu}$ and that of the intermediate $a \to bc$ decay is denoted by $\tilde{t}^{(l)\mu}$. The tensor $\tilde{T}^{(l)\mu}$ has the same definition as $\tilde{t}^{(l)\mu}$ in Ref.~\cite{covariant-tensors}.
\begin{table}[hbtp]
 \begin{center}
\begin{tabular}{| l c |}
\hline
 Decay chain& $S(p)$ \\
 \hline
$D^0[S]\to V_1V_2$ & $P^{\mu\nu}(D^0)\tilde{t}^{(1)}_{\nu}(V_1)\tilde{t}^{(1)}_\mu(V_2)$ \\
$D^0[P]\to V_1V_2$ & $\epsilon_{\mu\nu\lambda\sigma}p^\mu(D^0) \; \tilde{T}^{(1)\nu}(D^0)\tilde{t}^{(1)\lambda}(V_1) \; \tilde{t}^{(1)\sigma}(V_2) $ \\
$D^0[D]\to V_1V_2$ & $\tilde{T}^{(2)\mu\nu}(D^0)\tilde{t}^{(1)}_\mu(V_1)\tilde{t}^{(1)}_\nu(V_2)$ \\
$D^0\to AP_1, A[S] \to VP_2$ & $\tilde{T}^{(1)\mu}(D^0) \; P^{(1)}_{\mu\nu}(A) \; \tilde{t}^{(1)\nu}{(V)}$ \\
$D^0\to AP_1, A[D] \to VP_2$ & $\tilde{T}^{(1)\mu}(D^0) \; \tilde{t}^{(2)}_{\mu\nu}(A) \; \tilde{t}^{(1)\nu}{(V)}$ \\
$D^0\to AP_1, A\to SP_2$ & $\tilde{T}^{(1)\mu}(D^0)\tilde{t}^{(1)}_\mu{(A)}$ \\
$D^0\to VS$ & $\tilde{T}^{(1)\mu}(D^0)\tilde{t}^{(1)}_\mu{(V)}$ \\
$D^0\to V_1P_1, V_1 \to V_2P_2$ & $\epsilon_{\mu\nu\lambda\sigma}p^\mu_{V1}r^\nu_{V1}p^\lambda_{P1}r^\sigma_{V2}$ \\
$D^0\to PP_1,P\to VP_2$ & $p^\mu(P_2)\tilde{t}^{(1)}_\mu{(V)}$ \\
$D^0\to TS$ & $\tilde{T}^{(2)\mu\nu}(D^0)\tilde{t}^{(2)}_{\mu\nu}(T)$ \\
\hline
\end{tabular}
 \caption{The spin factor $S(p)$ for each decay chain. All operators, i.e.~$\tilde{t}$ and $\tilde{T}$, have the same definitions as in Ref.~\cite{covariant-tensors}. Scalar, pseudo-scalar, vector, axial-vector and tensor states are denoted by $S$, $P$, $V$, $A$ and $T$, respectively. The $[S]$, $[P]$ and $[D]$ denote the orbital angular-momentum quantum numbers $L$ = 0, 1 and 2, respectively.\label{table:spin_factors}}
\end{center}
\end{table}
\subsection{Fit results}
Using the method described in Sec.~\ref{sec:fitmethod}, we perform the fit in steps by adding resonances one by one. The statistical significance of the newly added resonance is calculated by considering the change in the log likelihood value and the change in the number of degrees of freedom. Based on the theoretical predictions and the observed invariant mass distributions, the process $D^0\to K^{*}(892)^+ K^{*}(892)^-$ is expected to be a dominant process in the overall amplitude. Hence, the magnitude and phase of its $S$-wave process are fixed to 1.0 and 0.0 as reference, respectively, while those of other processes are varied. The $P$-wave and $D$-wave for decay $D^0\to K^{*}(892)^+ K^{*}(892)^-$ are also kept in the fit, even though the final significance of the latter contribution is 3.5$\sigma$.  
From inspection of the invariant masses in the data sample, there is a clear signal of the $\eta(1475)$ meson around 1.5~GeV/$c^2$ in the $M_{K^-K^+\pi^0}$ projection and a signal for the $a_0(980)$ meson around 1.0~GeV/$c^2$ in the $M_{K^-K^+}$ projection. Therefore, both these resonances will also be included in the fit.
Then, we try other possible intermediate resonances, such as $K_1(1270)^+$, $K_1(1400)^+$, $\eta(1405)$, etc., by adding them one by one and retaining those that have statistical significance greater than five standard deviations. Amplitudes for $D^0\to K^{*}(892)^+ K^{*}(892)^-$, $D^0\to K_{1}(1400)^+(\to K^{*}(892)^+\pi^0) K^-$, $D^0\to f_1(1420)(\to K^{*}(892)K)\pi^0$, $D^0\to \eta(1475)(\to K^{*}(892)K)\pi^0$, $D^0\to \eta(1405)(\to a_0(980)\pi^0)\pi^0$, meet this criteria and are retained in the nominal fit.
A comprehensive list of the tested contributions with statistical significances less than 5$\sigma$ that are not included in the nominal fit is provided in Appendix~\ref{app:test}.

The fit fraction (FF) for each amplitude is defined as the ratio between the integral of its absolute square over the PHSP and the integral of the absolute square of the total amplitude. For the $n^{\rm{th}}$ amplitude, the FF is expressed as
\begin{eqnarray}\begin{aligned}
  {\rm FF}_{n} = \frac{\int\left|\rho_{n}e^{i\phi_{n}}\mathcal{A}_{n}\right|^{2} d\Phi_{4}}{\int \left|\mathcal{M}\right|^{2} d\Phi_{4}}\,.
  \label{Fit-Fraction-Definition}
\end{aligned}\end{eqnarray}
In practice, the FF is computed numerically using generator-level PHSP MC events. The discrete form of Eq.~(\ref{Fit-Fraction-Definition}) becomes 
\begin{eqnarray}\begin{aligned}
  {\rm FF}_{n} = \frac{\sum^{N_{\rm gen}} \left|\rho_{n}e^{i\phi_{n}}\mathcal{A}_{n}\right|^{2}}{\sum^{N_{\rm gen}} \left|\mathcal{M}\right|^{2}}\,, 
  \label{Fit-Fraction-Definition-MC}
\end{aligned}\end{eqnarray}
where $N_{\rm gen}$ is the number of PHSP MC events at the generator level. The interference fit fractions between two components is defined as
\begin{eqnarray}\begin{aligned}
 {\rm F_iF_j} = \frac{\int 2 \mathbb{R}\left(\rho_{i}e^{i\phi_{i}}\rho_{j}e^{-i\phi_{j}}\mathcal{A}_{i}\mathcal{A}_{j}^{\star}\right) d\Phi_{4}}{\int \left|\mathcal{M}\right|^{2} d\Phi_{4}}\,,
    \end{aligned}
\end{eqnarray}
and are presented in Appendix~\ref{app:int}.

For the longitudinal polarization fraction ($F_L$) measurement, the helicity amplitudes $H_+$, $H_0$ and $H_-$ are constructed independently. 
The detailed calculations are presented in Appendix~\ref{app:polar}. 
By means of the conversion of helicity coefficients and covariant tensor coefficients, the longitudinal polarization fraction $F_L$ can be defined as:

\begin{equation}
    F_L=\frac{\int|H_0|^2d\Phi}{\int|H_-+H_0+H_+|^2d\Phi}=\frac{\sum^{N_{\rm{gen}}}|H_0|^2}{\sum^{N_{\rm{gen}}}|H_-+H_0+H_+|^2}.
\end{equation}
The integral is achieved with generator-level PHSP events. In this work, we determine $F_L(D^0\to K^{*}(892)^+K^{*}(892)^-)=0.468\pm0.046_{\rm{stat.}}\pm0.011_{\rm{syst.}}$. 

\begin{figure}
\centering
\includegraphics[width=5cm,height=4cm]{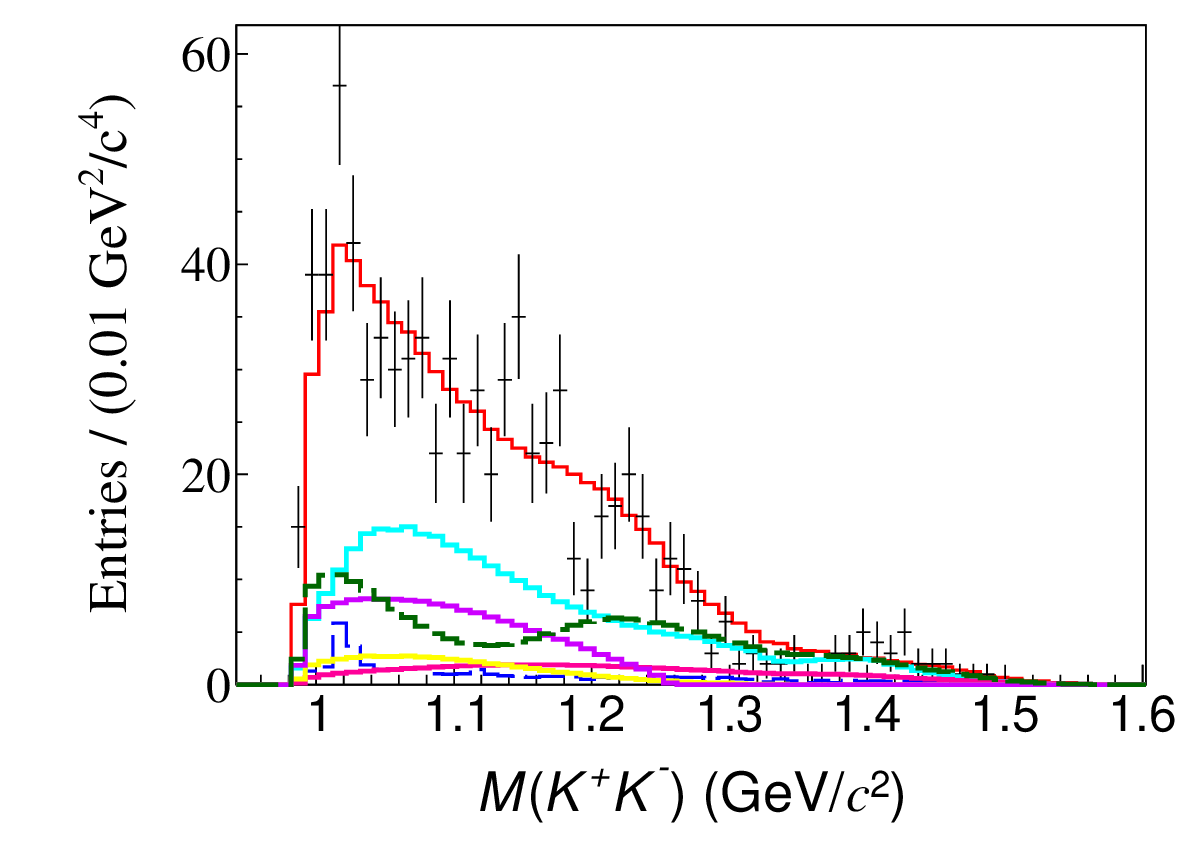}
\includegraphics[width=5cm,height=4cm]{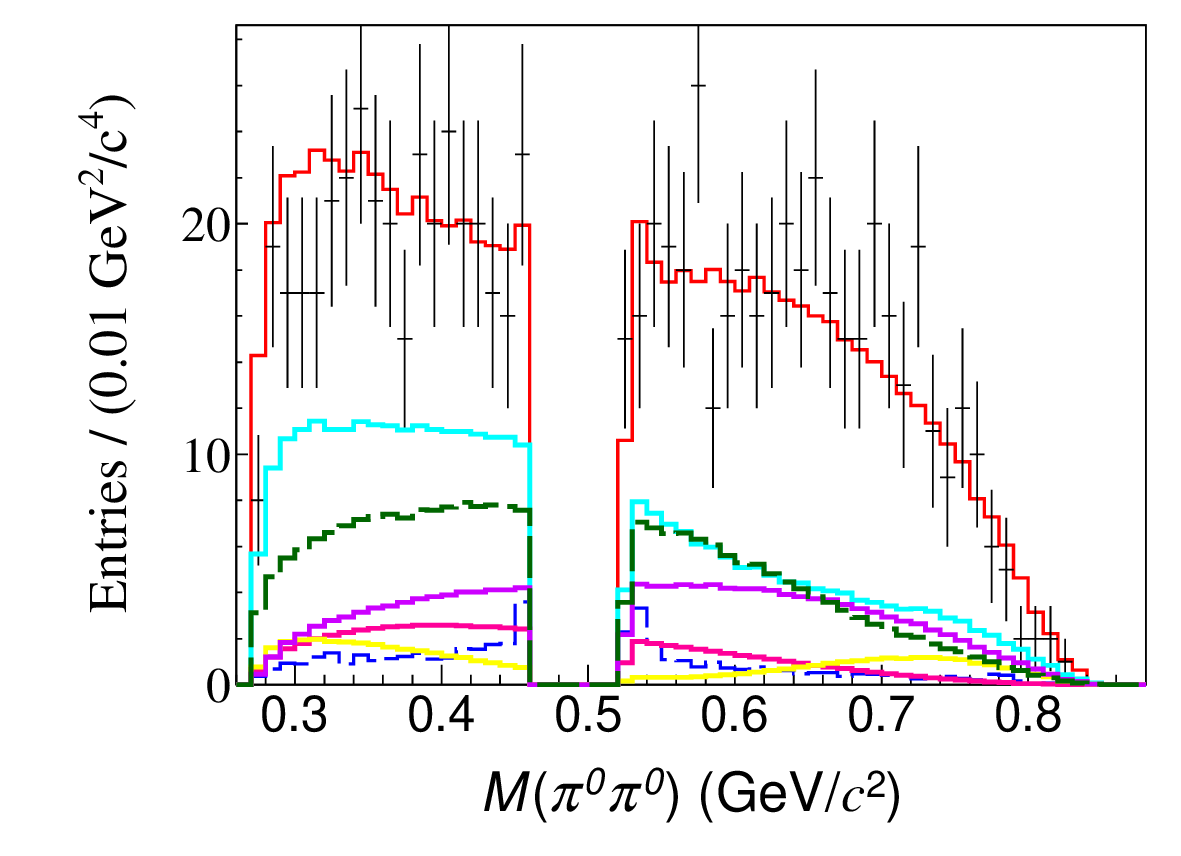}\\
\includegraphics[width=5cm,height=4cm]{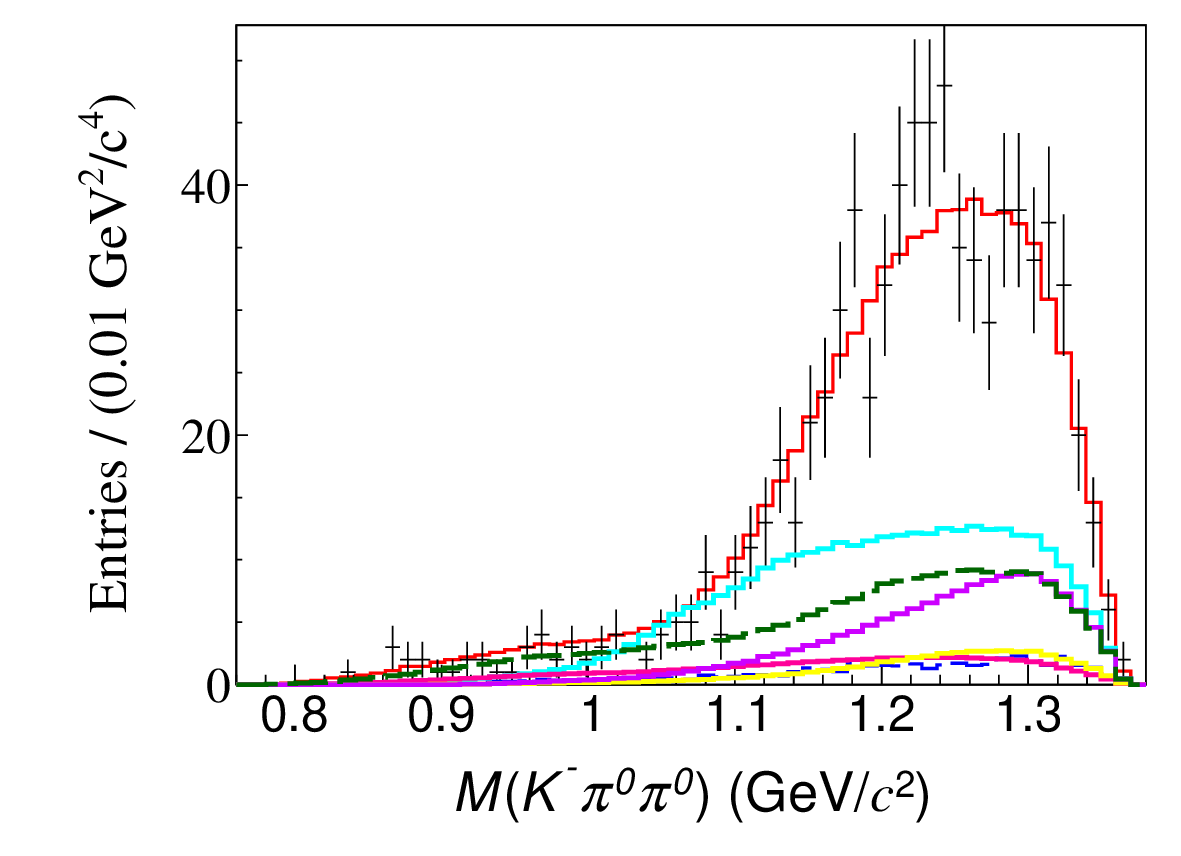}
\includegraphics[width=5cm,height=4cm]{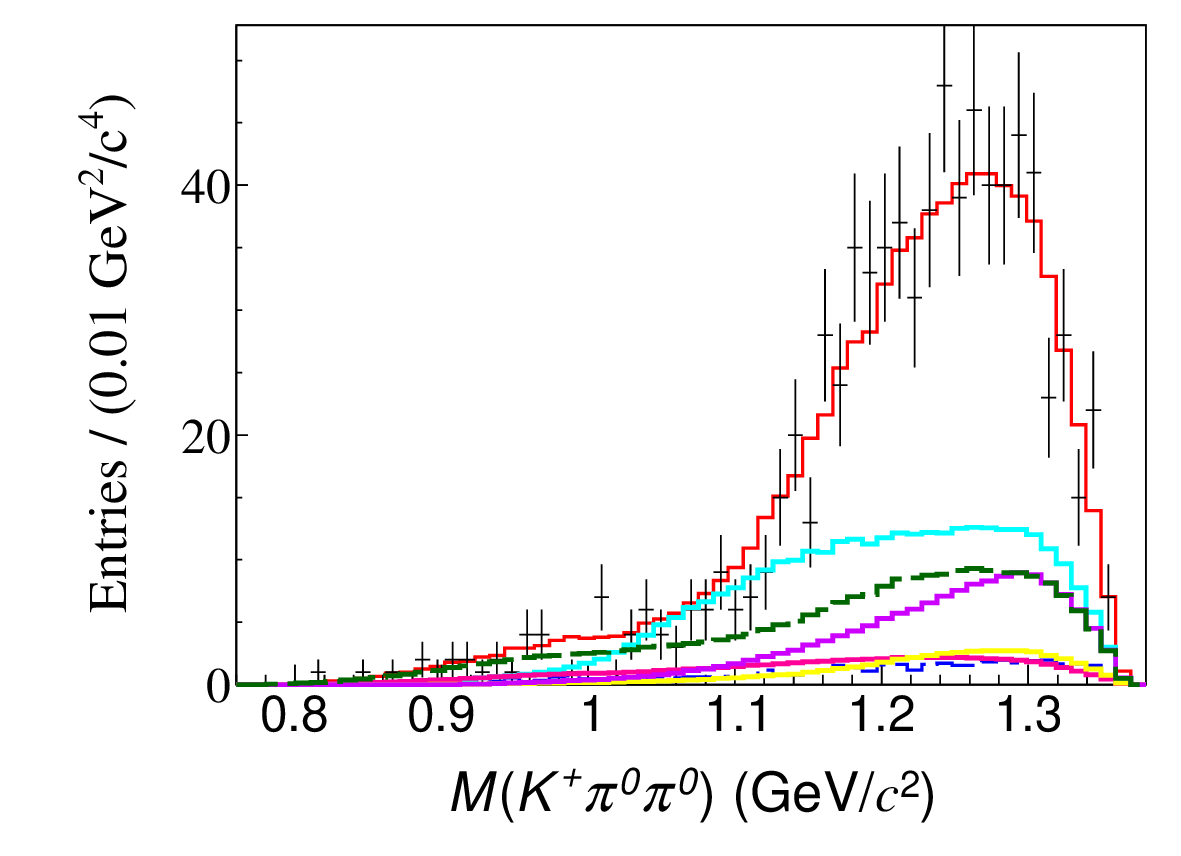}\\
\includegraphics[width=5cm,height=4cm]{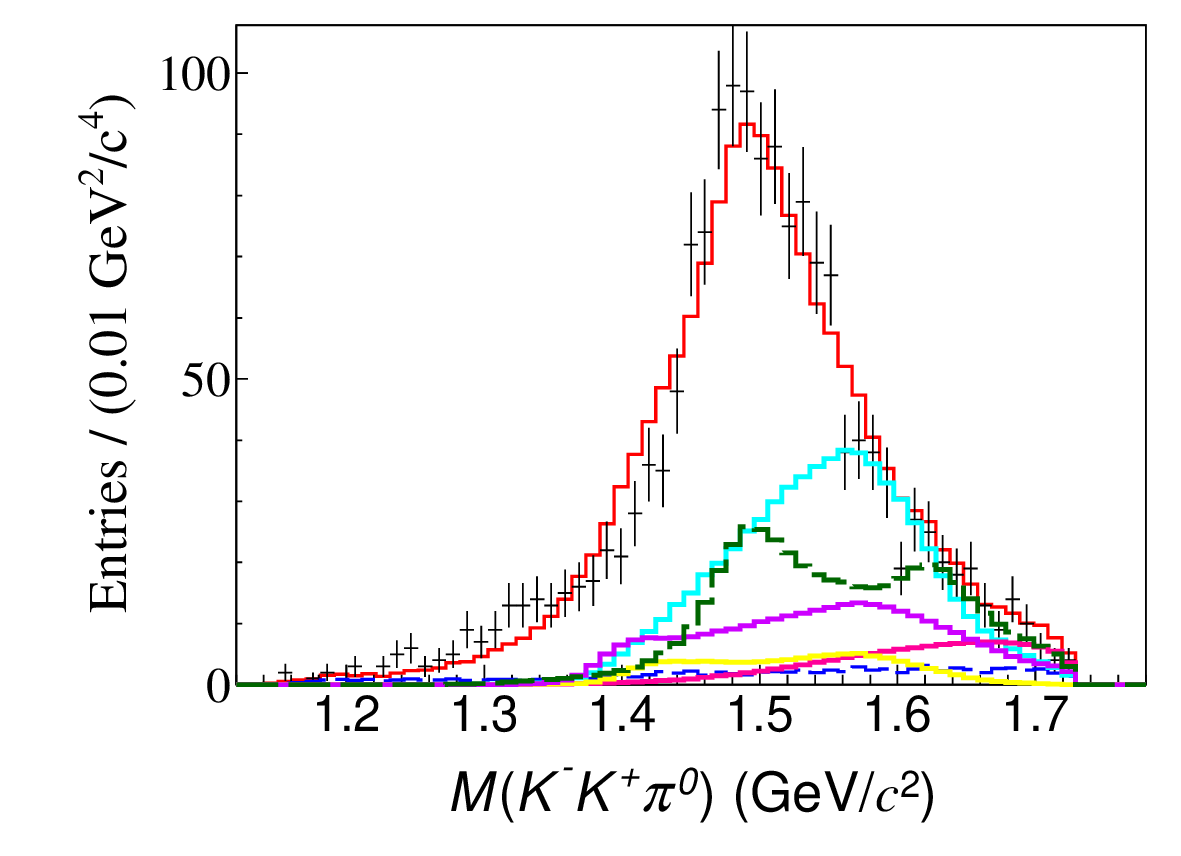}
\includegraphics[width=5cm,height=4cm]{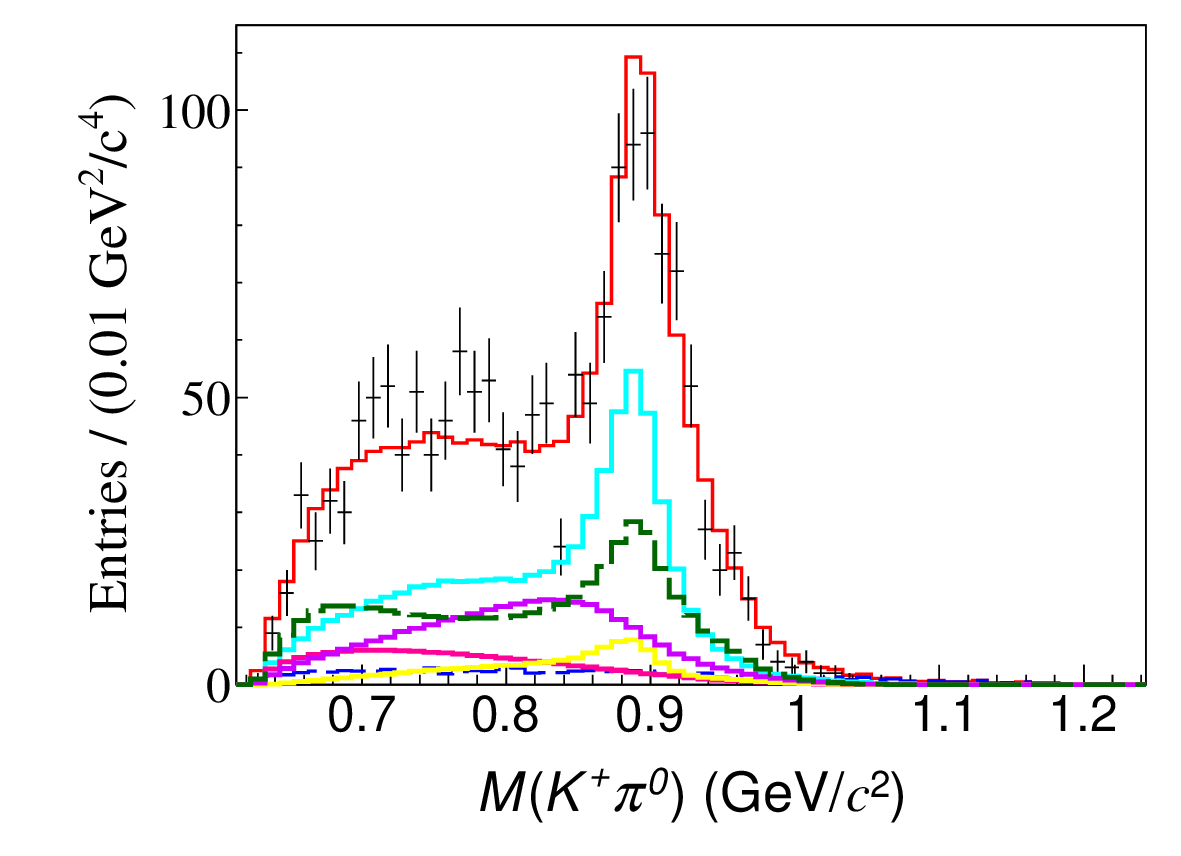}\\
\hspace{1cm}
\includegraphics[width=5cm,height=4cm]{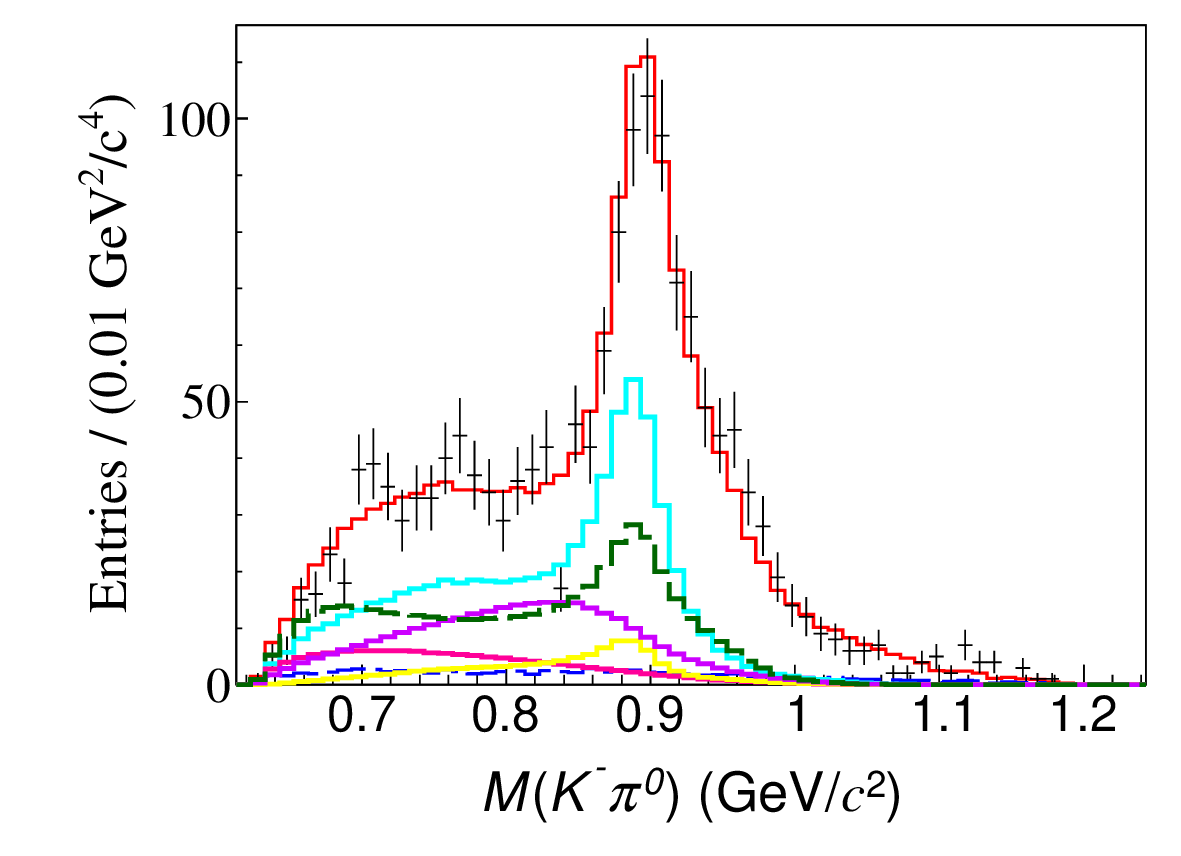}
\hspace{1cm}
\includegraphics[width=5cm,height=4cm]{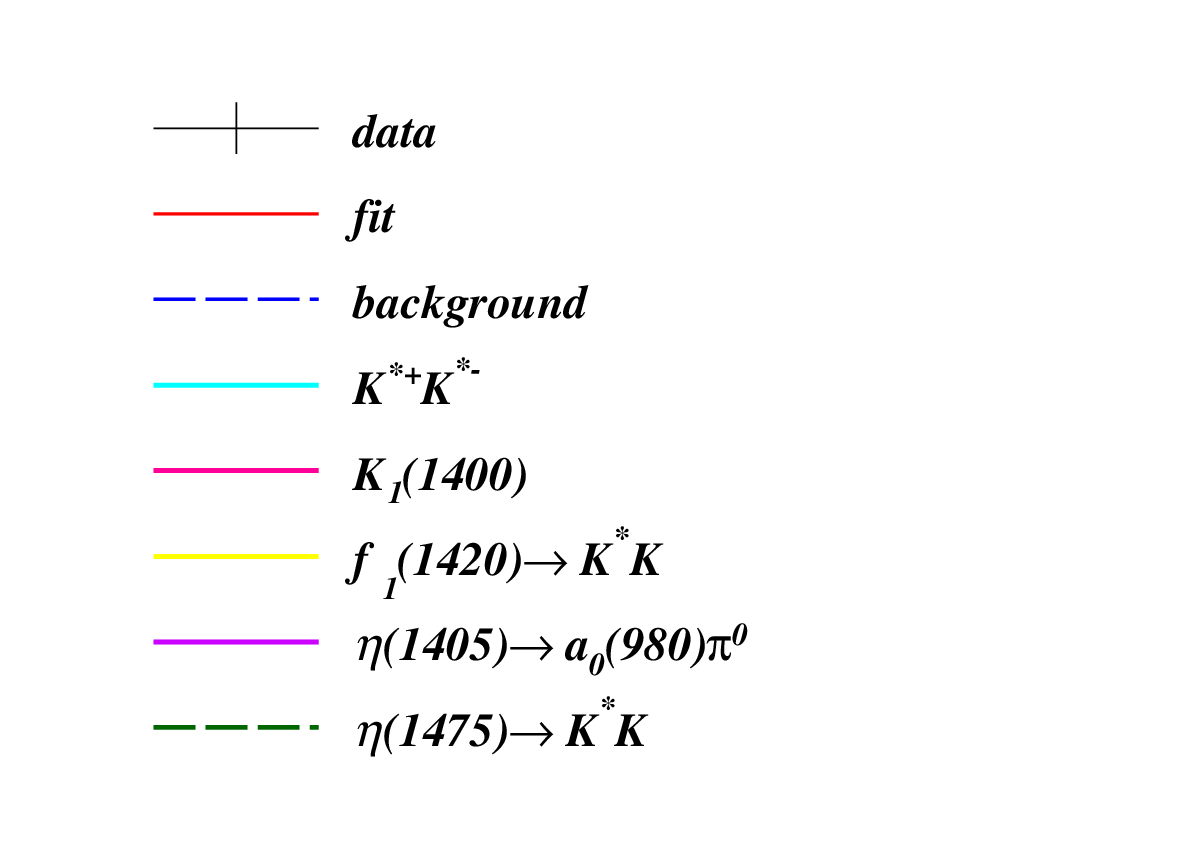}
 \caption{Projections of the data sample and the fit result on the invariant mass distributions. The data are represented by points with error bars, the fit results by colored lines, and the background by blue histograms. The combinations of two identical $\pi^0$s are added due to the exchange symmetry.}
    \label{fig:fitresult}
\end{figure}

It is impractical to analytically propagate the uncertainties from those of the magnitudes and phases to the FFs and $F_L$. Instead, the variables are randomly varied 500 times based on their covariance matrix obtained from the fit, and in each iteration, the FFs and $F_L$ are calculated to determine the statistical uncertainties. A Gaussian function is then used to fit the distribution. The width of this function is assigned as the uncertainty of the corresponding parameters. The magnitudes, phases， FFs and $F_L$ for different amplitudes are listed in Table~\ref{tab:signi}. The mass projections of the nominal fit are shown in Fig.~\ref{fig:fitresult}. One final item that can be determined from the amplitude model is the $CP-$even fraction of the decay $F_{+}$ as defined in Ref.~\cite{QCBF}. The central value is determined to be $F_{+}=0.89\pm.0.02$ where the uncertainty is statistical only. 
\begin{table}[htbp]
    \renewcommand{\arraystretch}{1.3}

    \centering
 \resizebox{\textwidth}{!}{
	\begin{tabular}{| l c r@{ $\pm$ }l@{ $\pm$ }c r@{ $\pm$ }c@{ $\pm$ }c c |}
	\hline
  Amplitude &Magnitude&\multicolumn{3}{c}{Phase $\phi_n$~(rad)} &\multicolumn{3}{c}{FF (\%)} &Significance ($\sigma$)\\
	\hline
    $D^0[S]\to K^{*}(892)^+ K^{*}(892)^-$
    &1 (fixed)&\multicolumn{3}{c}{{0.0 (fixed)}} &26.8 &3.5 &1.7 &$>$10\\
    $D^0[P]\to K^{*}(892)^+ K^{*}(892)^-$
    &$0.70\pm0.08$& 3.00 &0.16 &0.11 &8.0  &1.8 &0.9 &7.3\\
    $D^0[D]\to K^{*}(892)^+ K^{*}(892)^-$
    &$2.01\pm0.39$& 5.33 &0.20 &0.13 &4.4  &1.5 &0.5 &3.5\\
    $D^0\to K^{*}(892)^+ K^{*}(892)^-$
    &-&\multicolumn{3}{c}{-} &40.2  &4.2 &1.5 &-\\
    \hline
    $D^0\to K_{1}(1400)^+(\to K^{*}(892)^+ \pi^{0}) K^{-}$
    &$2.03\pm0.29$&1.17 &0.18 &0.12 &8.3 &2.0 &1.5 &6.8\\
    \hline
    $D^0\to f_1(1420)(\to K^{*}(892)^+ K^{-}) \pi^{0}$
    &$0.85\pm0.14$&1.38 &0.17 &0.24  &3.1  &0.8 &0.5 &-\\
    $D^0\to f_1(1420)(\to K^{*}(892)^- K^{+}) \pi^{0}$
    &$0.85\pm0.14$&1.38 &0.17 &0.24   &3.1  &0.8 &0.5 &-\\
    $D^0\to f_1(1420) \pi^{0}$
   &-&\multicolumn{3}{c}{-} &8.1 &2.1 &1.1 &7.7\\   
  \hline
    $D^0\to \eta(1405)(\to  a_{0}(980) \pi^{0})\pi^{0}$
    &$1.66\pm0.18$&2.25 &0.26  &0.30    &19.0 &2.9 &1.9 &7.8 \\
    \hline
    $D^0\to \eta(1475)(\to K^{*}(892)^-K^+) \pi^0$
    &$2.06\pm0.20$&1.20 &0.15  &0.20  &14.8  &1.7 &1.2 &-\\
    $D^0\to \eta(1475)(\to K^{*}(892)^+K^-) \pi^0$
    &$2.06\pm0.20$&1.20 &0.15  &0.20 &14.8  &1.7 &1.2 &-\\
    $D^0\to \eta(1475)\pi^0$
     &-&\multicolumn{3}{c}{-}  &33.7 &3.9 &2.4 &$>$10 \\
    \hline
    Total&\multicolumn{4}{c}{ }&\multicolumn{3}{c}{102.3} & \\
    \hline
    \end{tabular}}
  \caption{Magnitude, phases, FFs, and statistical significances for different amplitudes in $D^0\to K^+K^-\pi^0\pi^0$. Groups of related amplitudes are separated by horizontal lines and the last row of each group gives the total fit fraction of the above components with interferences considered. The first and second uncertainties for the phases and FFs are statistical and systematic, respectively. The letters in bracket represent relative orbital angular momentum between resonances. }	  
	\label{tab:signi}
\end{table}

\subsection{Systematic uncertainties in the amplitude analysis}
\label{sec:PWA-Sys}
The systematic uncertainties in the amplitude analysis are described below and summarized in Table~\ref{tab:pwa_sys}. 
\begin{itemize}

\item[\uppercase\expandafter{\romannumeral1}] 
Parameters in propagators: The masses and widths of resonances are varied to their $\pm 1\sigma$ boundaries as tabulated in~\cite{PDG}, with two dedicated refits performed to evaluate the corresponding systematic uncertainties.
The $K\pi$ $S$-wave model is modified according to the parameters in Table~\ref{tab:babar}. For each of the $K\pi$ $S$-wave model formalisms, two independent fits are performed by setting the parameters to their upper and lower $1\sigma$ limits. 
The systematic uncertainty for each parameter is determined by its maximum deviation between these alternative and nominal results. The total amplitude model uncertainty is calculated by combining all individual contributions in quadrature.

\item[\uppercase\expandafter{\romannumeral2}]
Effective radius: The estimation of the systematic uncertainty associated with the $R$ parameters in the Blatt-Weisskopf factors is performed by repeating the fit procedure twice, varying the effective radius of the intermediate states and that of the  $D^0$ meson by $\pm 1~({\rm GeV}/c)^{-1}$ in separate fits.

\item[\uppercase\expandafter{\romannumeral3}]  
Background: When analyzing the impact of background on the amplitude model, there are two primary sources of influence: the magnitude and shape of the background.  
The background size is associated with the signal purity ($w_{\rm{sig}}$) in Eq.~(\ref{likelihood3}). 
The systematic uncertainty from the $w_{\rm{sig}}$ parameter is evaluated by performing two additional fits with the parameter fixed at its $\pm 1 \sigma $ boundaries. The corresponding systematic uncertainty for each observable is then taken as the maximum deviation between these boundary fits and the nominal result. The background shape is related to the background function ($B(p_j)$) in Eq.~(\ref{likelihood3}). Two alternative background samples are used to determine the background shape, where the relative fractions of background processes from direct $e^+e^-\to \bar{q} q$ are varied up and down by the uncertainty of the known cross section~\cite{Luminosity}. The maximum relative change of the fit result is assigned as the corresponding systematic uncertainty. The square root of the quadratic sum of these two uncertainties is taken as the uncertainty.

\item[\uppercase\expandafter{\romannumeral4}] 
Experimental effects: The systematic uncertainty from the $\gamma_{\epsilon}$ factor in Eq.~(\ref{pwa:gamma}), which corrects for data-MC differences in tracking, PID, as well as $\pi^0$ reconstruction efficiencies, is evaluated by performing the fit after varying the weights according to their uncertainties.  

\item[\uppercase\expandafter{\romannumeral5}] 
Fit bias: To study the possible bias from the fit procedure, an ensemble of 600 signal MC samples is generated according to the results of the amplitude analysis. The fit procedure is repeated for each signal MC sample, and the pull distributions of the amplitude results are fitted by a Gaussian. The fit results show that no obvious biases and under- or over-estimations on statistical uncertainties are observed. Therefore, this systematic uncertainty is neglected. 

\item[\uppercase\expandafter{\romannumeral6}]
Insignificant amplitudes: To achieve more reliable estimation of the model systematic effects, we have checked the impact of the intermediate processes with statistical significances less than 5$\sigma$ in App.~\ref{app:test}. The model including $D^0\to K_1(1400)^-K^+, K_1(1400)^-\to K^{*}(892)^-\pi^0$ has the largest variation with the nominal fit. The variation of the phases and FFs from the nominal result is taken as the corresponding systematic uncertainty.

\end{itemize}

 \begin{table}[htbp]
             \label{tab:pwa_sys}
\caption{Systematic uncertainties on the phases, FFs, $F_p$ and $F_+$ for different amplitudes in units of the corresponding statistical uncertainties. (I) Parameters in propagators, (II) R value, (III) experiment effect, (IV) background, (V) Insignificant amplitudes.}
  \centering
  \scalebox{0.87}{
 \begin{tabular}{|lccccccc|}   
    \hline
    	\   &\multicolumn{7}{c|}{Source}   \\
     \hline
    Amplitude  & & I &II&III&IV&V&Total \\
    \hline	
    $D^0[S]\to K^{*}(892)^+ K^{*}(892)^-$
    & FF   & 0.41   &0.16&  0.08     &  0.03     &0.21    &0.49        \\
    \multirow{2}{*}{$D^0[P]\to  K^{*}(892)^+ K^{*}(892)^-$} 
    & FF   &0.51    &0.13&  0.02     & 0.12       &0.13    & 0.56       \\
    &$\phi$&0.50    &0.10&  0.31     & 0.30       &0.07    &0.70        \\
    \multirow{2}{*}{$D^0[D]\to  K^{*}(892)^+ K^{*}(892)^-$}
    & FF   &0.22    &0.04&  0.24     & 0.12      &0.18    &0.36       \\
    &$\phi$&0.49    &0.10&  0.22     & 0.32      &0.01   & 0.64       \\
    $D^0\to K^{*}(892)^+ K^{*}(892)^-$
    & FF   & 0.31   &0.02&  0.14     &  0.06     &0.03    &0.35        \\   
    \multirow{2}{*}{$D^0\to K_1(1400)^+ K^{-}, K_1(1400)^+\to K^{*}(892)^+ \pi^{0}$}
    &  FF  &       0.65  &0.23     &     0.30  &      0.09     &0.03 &0.76  \\
    & $\phi$  &       0.33 & 0.10    &   0.20    &   0.49      &0.22 &0.67        \\
    \multirow{2}{*}{$D^0\to f_1(1420)\pi^{0}, f_1(1420)\to K^{*}(892)^+ K^{-}$}
    &  FF  &        0.59    &   0.03  & 0.21      & 0.09          &0.12 &   0.65  \\
    & $\phi$ &       1.37  &    0.03 &      0.22 &  0.03         &0.04& 1.39       \\
    \multirow{2}{*}{$D^0\to f_1(1420)\pi^{0}, f_1(1420)\to K^{*}(892)^- K^{+}$}
    &   FF &        0.59  &     0.03&   0.20     &  0.10         &0.12 &   0.65     \\
    & $\phi$  &       1.37 & 0.03    &  0.22     &   0.03    &0.04& 1.39       \\
    $D^0\to f_1(1420)\pi^{0}, f_1(1420)\to K^{*}(892) K$
    &   FF &        0.46  &     0.01&   0.22     &  0.08         &0.04 &   0.52\\
    \multirow{2}{*}{$D^0\to \eta(1405)\pi^{0}, \eta \to a_{0}(980) \pi^{0}$}
    &  FF  &          0.61 &    0.18 &     0.12  &  0.20        &0.10  &      0.69  \\
    & $\phi$  &       0.39&     2.12  &       0.45&  0.13     &0.18&2.22        \\
    \multirow{2}{*}{$D^0\to \eta(1475)\pi^{0}, \eta(1475)\to K^{*}(892)^+ K^{-}$}
    &   FF &    0.34      &     0.48    &  0.31     &   0.33       &0.11   &       0.75 \\
    & $\phi$ &      1.18  &         0.24&     0.44  &       0.17  &0.20  &       1.35 \\
    \multirow{2}{*}{$D^0\to \eta(1475)\pi^{0}, \eta(1475)\to K^{*}(892)^- K^{+}$}
    &  FF  &    0.34      & 0.48        & 0.32      &   0.33            &0.11 &   0.75     \\
    &  $\phi$ &    1.18    &     0.24        & 0.44         &0.17       &0.20  &1.35        \\
    $D^0\to \eta(1475)\pi^{0}, \eta(1475)\to K^{*}(892) K$
     &  FF  &    0.57      & 0.02        & 0.18      &0.14    &0.14 &   0.63     \\
    $F_L(D^0\to K^{*}(892)^+ K^{*}(892)^-)$ 
    &          &0.07 &0.08 &0.03 &0.11  &0.09 &0.19\\ 
    $F_+$ &          &0.63 &0.24 &0.14 &0.00 &0.31 &0.76\\
    
    \hline

\end{tabular}}
\end{table}

\section{Branching fraction measurement}
The BF of  $D^0 \to K^+K^-\pi^0\pi^0$ is measured with the DT technique, applying the same tag modes as those utilized in the amplitude analysis.
The selection criteria follow those discussed in Sec.~\ref{ST-selection}, and are identical to the criteria used in the amplitude analysis.

For a given ST mode, the following relations are established~\cite{QCBF}:
\begin{equation}
  N_{\text{tag}}^{\text{ST}} = 2N_{D^{0} \bar D^{0}}\cdot \mathcal{B}_{\text{tag}}\cdot\epsilon_{\text{tag}}^{\text{ST}}\cdot(1+y_D^2)\cdot(1+r_{\rm tag}^2-2r_{\rm tag} R_{\rm tag} y_D \cos \delta_{\rm tag})\,, \label{eq-ST}
\end{equation}
\begin{equation}
\begin{split}
N_{\text{tag,sig}}^{\text{DT}} = & 2N_{D^{0} \bar D^{0}}\cdot \mathcal{B}_{\rm sub}\cdot \mathcal{B}_{\text{tag}}\cdot\mathcal{B}_{\text{sig}}\cdot \epsilon_{\text{tag,sig}}^{\text{DT}} \\
& \cdot(1+y_D^2)\cdot[1+r_{\rm tag}^2-2r_{\rm tag} R_{\rm tag} \cos \delta_{\rm tag}(2F_{+}^{\rm sig}-1)]\,,
\end{split}
\label{eq-DT}
\end{equation}
where $N_{\text{tag}}^{\text{ST}}$ is the ST yield for a specific tag mode, $N_{D^{0} \bar D^{0}}$ is the total number of $D^{0} \bar D^{0}$ pairs produced from $e^{+}e^{-}$ collisions, $\mathcal{B}_{\text{tag}}$ is the BF of the tag mode, and $\epsilon_{\text{tag}}^{\text{ST}}$ is the ST efficiency for the tag mode. The DT yield is denoted $N_{\text{tag,sig}}^{\text{DT}}$, $\mathcal{B}_{\text{sig}}$ is the BF of the signal mode, and $\epsilon_{\text{tag,sig}}^{\text{DT}}$ is the efficiency for simultaneously reconstructing the signal and a specific tag mode. To account for the reconstruction of the signal through subsequent decays, the factor $\mathcal{B}_{\text{sub}} = \mathcal{B}^2(\pi^0 \to \gamma\gamma)$ is introduced. Additionally, $y_D$ is the $D^0-\bar D^0$ mixing parameter.
For the $D^0 \to K^+K^-\pi^0\pi^0$ decay, $F_+ = 0.89\pm0.02\pm0.01$ from the amplitude model determined here. The parameters $r$, $R$, and $\delta$ are introduced to account for quantum-correlation effects due to the tags not being pure flavour specific, and their values for the three tag modes are listed in Table~\ref{tab:qcc}. Combining the two equations above and ignoring the term $2r_{\rm tag} R_{\rm tag} y_D \cos \delta_{\rm tag}$, the absolute BF of  $D^0 \to K^+K^-\pi^0\pi^0$ is determined by
\begin{equation}
  \mathcal{B}_{\text{sig}} = \frac{N_{\text{tag,sig}}^{\text{DT}}}{\begin{matrix} \mathcal{B}_{\rm sub}\cdot N_{\text{tag}}^{\text{ST}}\cdot \epsilon^{\text{DT}}_{\text{tag,sig}}/\epsilon_{\text{tag}}^{\text{ST}}\cdot[1-\frac{2r_{\rm tag} R_{\rm tag} \cos \delta_{\rm tag}}{1+r_{\rm tag}^2}(2F_{+}^{\rm sig}-1)]\end{matrix}}\,.\label{BR-formula}
\end{equation} 

\begin{table}[hbtp]
  \begin{center}
  \begin{tabular}{|l r@{ $\pm$ }l r@{ $\pm$ }l r@{ $\pm$ }l |}
      \hline
      Tag mode &\multicolumn{2}{c}{$r~(\%)$}  &\multicolumn{2}{c}{$R$}  &\multicolumn{2}{c|}{$\delta~(^{\circ})$} \\ 
      \hline
      $\bar {D}^0\to K^+\pi^-$	    &\multicolumn{2}{c}{$5.855^{+0.009}_{-0.010}$\cite{QCC1}} 
      &\multicolumn{2}{c}{1} 
      &\multicolumn{2}{c|}{$191.4 \pm 2.4$\cite{QCC1}}   \\
      $\bar {D}^0\to K^+\pi^-\pi^0$   &4.41 &0.11\cite{QCC2} 
      &0.79 &0.04\cite{QCC2}
      &196 &11\cite{QCC2} \\
      $\bar {D}^0\to K^+\pi^-\pi^-\pi^+$ &5.50 &0.07\cite{QCC2} 
      &\multicolumn{2}{c}{$0.44^{+0.09}_{-0.10}$\cite{QCC2}}
      &\multicolumn{2}{c|}{$161^{+28}_{-18}$\cite{QCC2}}\\
      \hline
    \end{tabular}
    \caption{ The input values of $r$, $R$ and $\delta$ for the three tag modes.}
    \label{tab:qcc}
  \end{center}
\end{table}

\begin{figure}[htbp]
  \centering
  \includegraphics[width=0.45\textwidth]{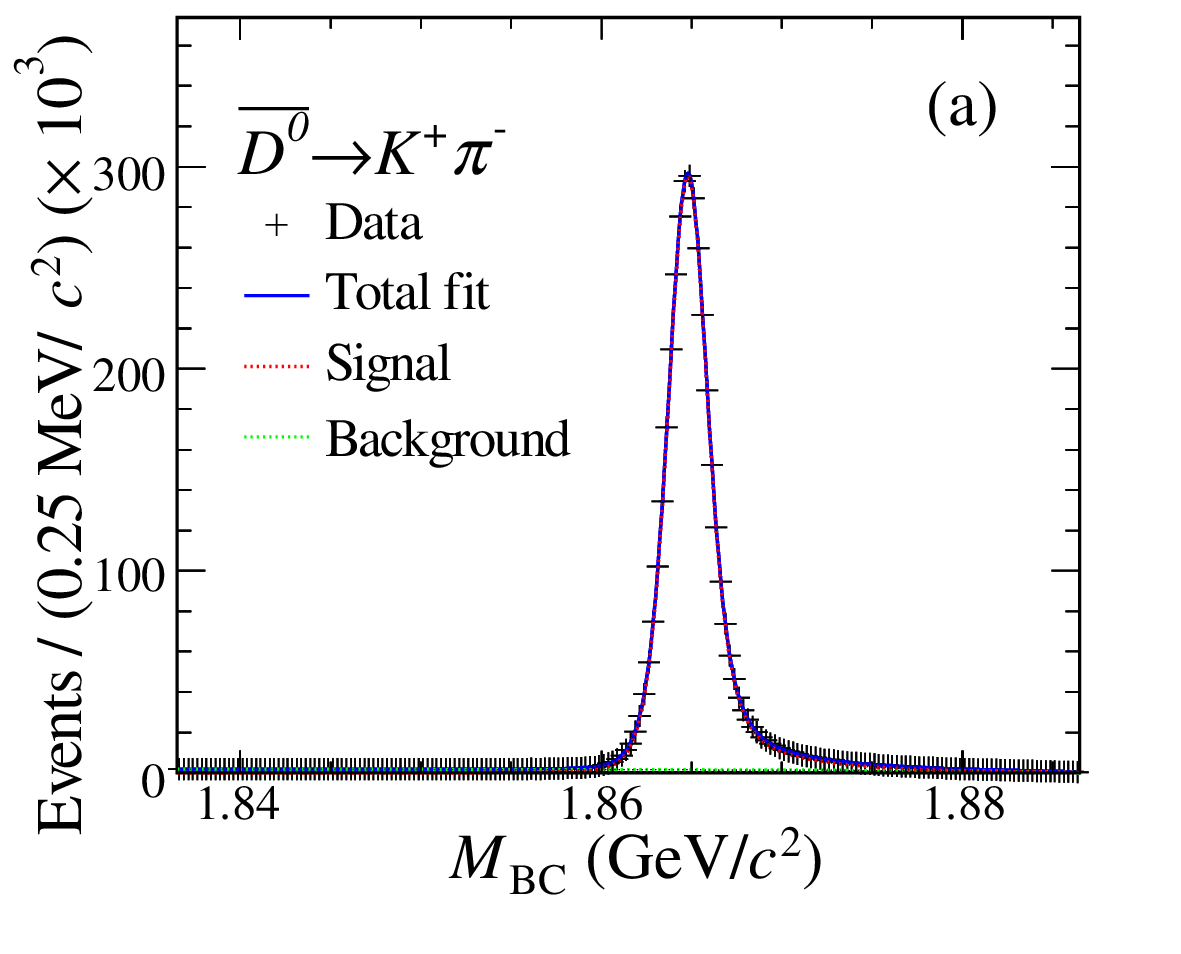}
  \includegraphics[width=0.45\textwidth]{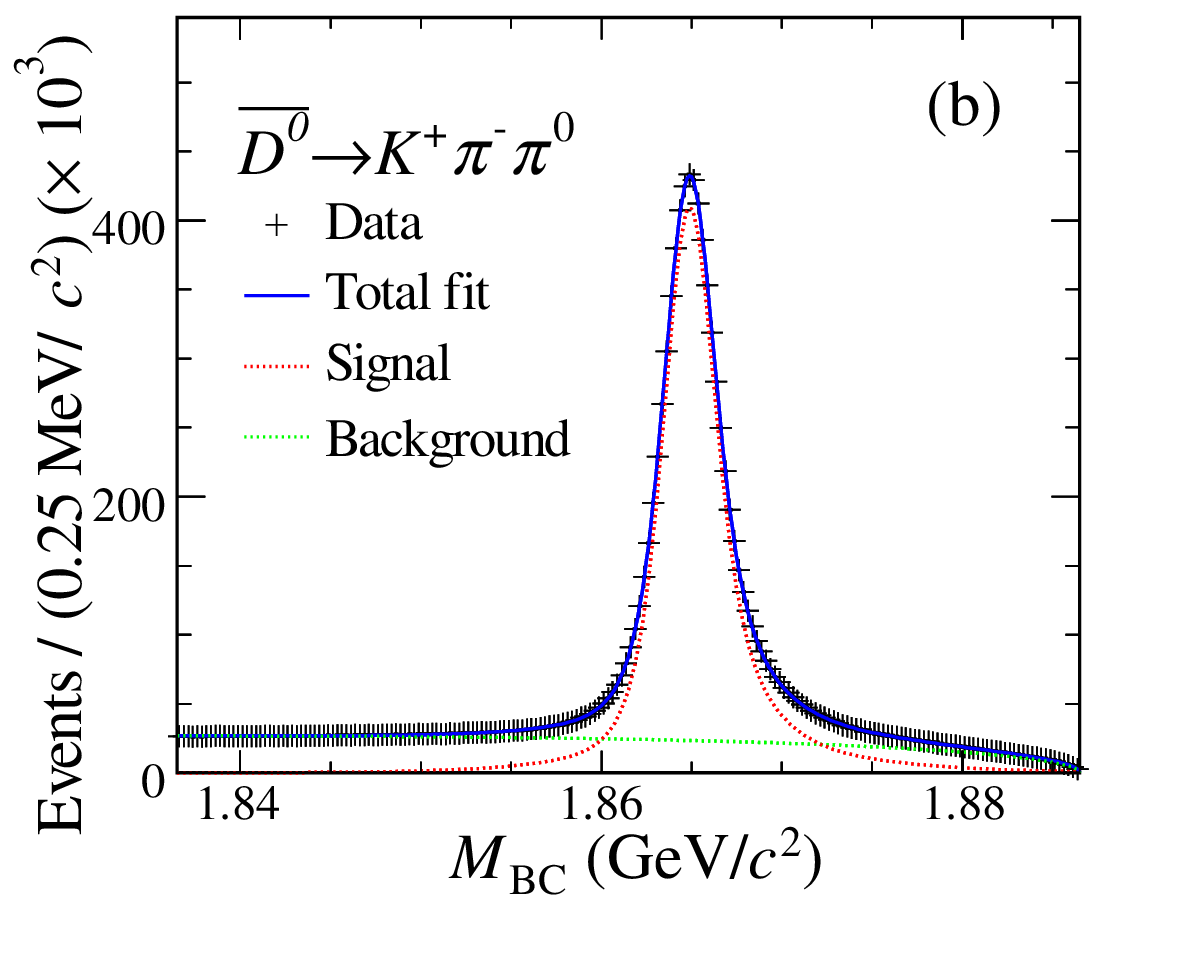}
  \includegraphics[width=0.45\textwidth]{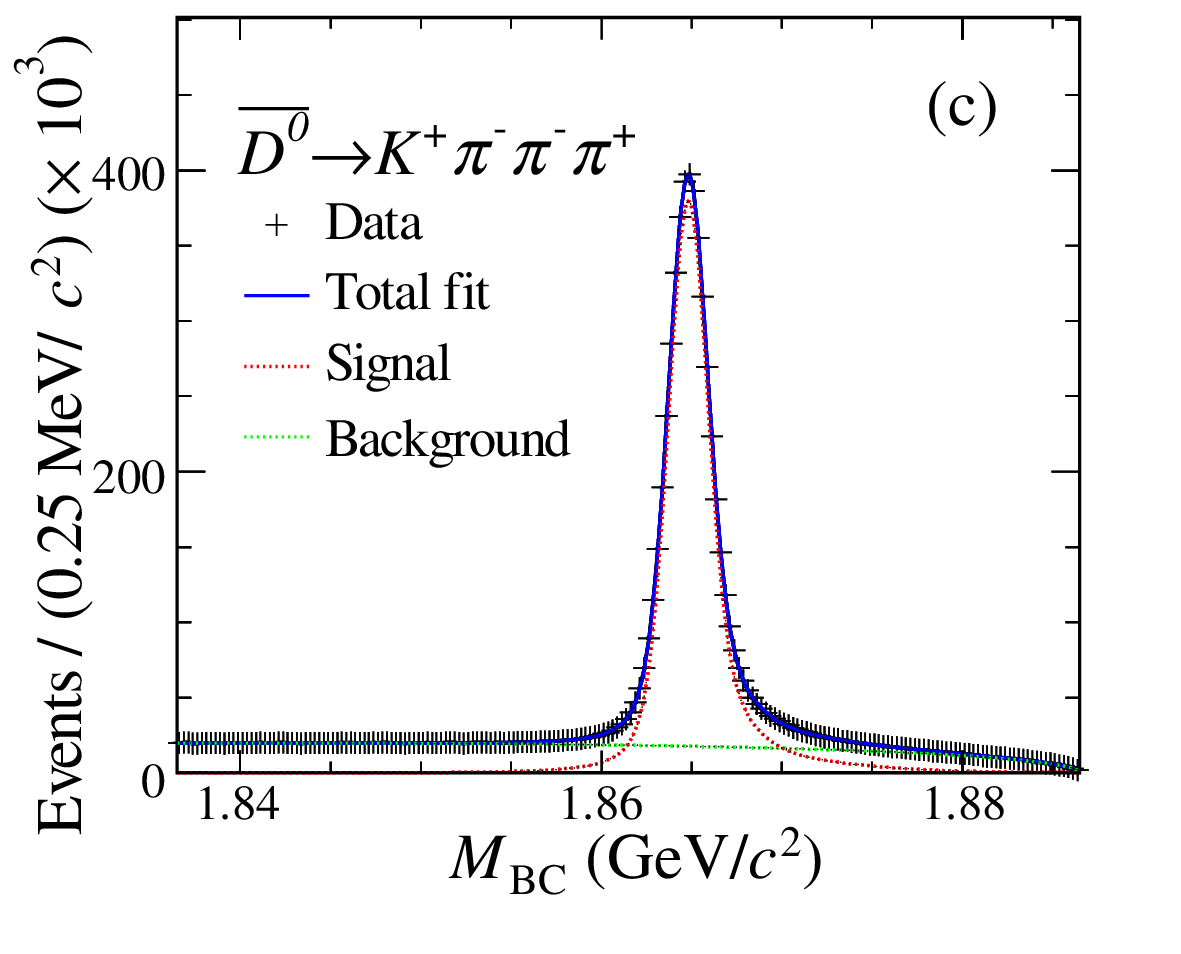}
  \caption{Fits to the $M_{\rm BC}$ distributions of the ST candidates for $\bar D^0 \to K^+ \pi^-$~(a), $\bar D^0 \to K^+ \pi^- \pi^0$~(b) and $\bar D^0 \to K^+ \pi^- \pi^- \pi^+ $~(c). The points with error bars are data. The blue curves are the fit projections. The dotted red curves are the signal. The dotted green curves are the fitted combinatorial background shapes.}
  \label{fig:ST_yield}
\end{figure}

The value of $N_{\text{tag}}^{\text{ST}}$ is obtained from a one-dimensional (1D) binned fit to the $M_{\rm BC}$ distribution after subtracting the peaking background, as shown in Fig.~\ref{fig:ST_yield}. 
The signal shape is modeled by an MC-simulated shape convolved with a double-Gaussian function describing the resolution difference between data and MC simulation, and the background shape is described by an ARGUS function~\cite{ARGUS}. The corresponding $\epsilon_{\text{tag}}^{\text{ST}}$ is estimated with the inclusive MC sample, where the peaking backgrounds have been removed from the samples.

The total DT yield over the entire $M_{\rm BC}$ region is determined to be $N_{\text{tag,sig}}^{\text{DT}} = 814 \pm 31$ via a 2D fit to the distribution of $M_{\rm BC}^{\rm tag}$ versus $M_{\rm BC}^{\rm sig}$, which is the same as shown in Fig.~\ref{fig:2dfit}. The efficiency, $\epsilon^{\text{DT}}_{\text{tag,sig}}$, is determined with the signal MC sample in which the $D^0 \to K^+K^-\pi^0\pi^0$ events are generated according to the result of the amplitude analysis.
The values of these efficiencies are summarized in Table~\ref{ST-eff}.

\begin{table}[htbp]
    \begin{center}
      \begin{tabular}{|lccc|}
        \hline
        Tag mode         &$N^{\text{ST}}_{\text{tag}} (\times 10^{3})$  & $\epsilon^{\text{ST}}_{\text{tag}}(\%)$  &$\epsilon^{\text{DT}}_{\text{tag,sig}}(\%)$ \\
        \hline
	$\bar {D}^0\to K^+\pi^-$              &$3820.8 \pm 2.0$  &$ 66.67 \pm 0.01$  &$~5.31 \pm 0.01$   \\  
	$\bar {D}^0\to K^+\pi^-\pi^0$         &$7926.5 \pm 3.3$  &$ 37.92 \pm 0.01$  &$~2.57 \pm 0.01$   \\ 
	$\bar {D}^0\to K^+\pi^-\pi^-\pi^+$    &$5140.1 \pm 2.6$  &$ 42.20 \pm 0.01$  &$~2.12 \pm 0.01$   \\  
        \hline
      \end{tabular}
    \end{center}
\caption{The ST yields ($N^{\text{ST}}_{\text{tag}}$), ST efficiency ($\epsilon^{\text{ST}}_{\text{tag}}$) and DT efficiency ($\epsilon^{\text{DT}}_{\text{tag,sig}}$). The efficiencies do not include the BF for $\pi^0 \to \gamma\gamma$. The uncertainties are statistical only.}
\label{ST-eff}
\end{table}

The systematic uncertainties for the BF measurement are described below and summarized in Table~\ref{BF-Sys}. The total systematic uncertainty is obtained by adding all individual contributions in quadrature. 

\begin{itemize}
\item ST $\bar D^{0}$ candidates:
The uncertainty in the yield of ST $\bar D^0$ mesons is assigned to be 0.3\% by varying the signal shape, background shape, and the parameters of the Gaussian in the fit. 

\item Tracking and PID:
The tracking and PID efficiencies of $K^{\pm}$ are investigated with the DT hadronic $D\bar{D}$ events of the decays $D^0\to K^-\pi^+$, $K^-\pi^+\pi^0$, $K^-\pi^+\pi^+\pi^-$ versus $\bar{D}^0\to K^+\pi^-$, $K^+\pi^-\pi^0$, $K^+\pi^-\pi^-\pi^+$, and $D^+\to K^-\pi^+\pi^+$ versus $D^- \to K^+\pi^-\pi^-$.
The data-MC efficiency ratios for $K$ tracking and PID are found to be 0.997 $\pm$ 0.001 and 1.001$\pm$0.001, respectively. After applying this correction factor to the MC efficiency for each $K$, the statistical uncertainty of the correction factor is propagated as the systematic uncertainty of each $K$. This results in a systematic uncertainty of 0.2\%  on the overall $K$ tracking or PID. 

\item $\pi^{0}$ reconstruction:
The data-MC efficiency ratio for each $\pi^0$ reconstruction is $0.974 \pm 0.002$, which is measured with the samples of $D^0 \to K^-\pi^+ \pi^0$ versus $\bar{D}^0 \to K^+\pi^-$ and $\bar{D}^0 \to K^+\pi^-\pi^- \pi^+$ hadronic decays. After applying this factor to correct the efficiency of each $\pi^0$ reconstruction, the statistical uncertainty of the correction factor is propagated to the systematic uncertainty. Since the analysis involves two $\pi^0$  mesons, this results in a total systematic uncertainty of 0.4\% on the overall reconstruction efficiency.

\item MC sample size:
The uncertainty of limited MC sample size is obtained by $\sqrt{ \begin{matrix} \sum_{i} (f_{i}\frac{\delta_{\epsilon_{i}}}{\epsilon_{i}}\end{matrix})^2}$, where $f_{i}$ is the tag yield fraction, $\epsilon_{i}$ is the signal efficiency and $\delta_{\epsilon_i}$ is the uncertainty of signal efficiency of tag mode $i$. The corresponding uncertainty is 0.1\%.

\item Amplitude model:
The uncertainty from the amplitude model is determined by varying the amplitude model parameters based on their error matrix 600 times. A Gaussian function is used to fit the distribution of 600 DT efficiencies, and the fitted width divided by the mean value is taken as an uncertainty, which is 0.3\%. The uncertainty associated with the amplitude model is estimated by varying the fitted parameters based on the covariance matrix. 

\item Quoted ${\mathcal B}(\pi^0\to\gamma\gamma)$:
In this measurement, the BF of the daughter particles are quoted from the PDG~\cite{PDG}, which is $\mathcal{B}(\pi^0 \to \gamma \gamma)=(98.823\pm 0.034)\%$. 
This gives an associated systematic uncertainty of 0.1\% for the total BF.   

\item 2D fit:
  The signal and background shapes are potential sources of uncertainty from the 2D fit. The mean and width of the convolved Gaussian function are varied by $\pm1\sigma$ for the signal shape and the $e^+e^-\to q\bar q$ component in the inclusive MC sample are varied by the statistical uncertainty of the known cross section.~\cite{Luminosity} for the background shape. The quadratic sum of the relative BF changes, 0.3\%, is assigned as the systematic uncertainty for the 2D fit.

\item $\theta_{D^0\bar{D}^0}$:
The efficiencies of the $D^0\bar{D}^0$ opening angle requirement is
studied by using the DT events of $D^0 \to K^- \pi^+ \pi^0\pi^0$ tagged by the same tag modes in our analysis. The difference in accepted efficiencies between the data and MC simulations, which is $0.5\%$ for this decay, is assigned as the associated systematic uncertainty.

\item $\Delta E_{\rm sig}$ requirement:
The possible difference between data and MC simulation is accounted for by examining the $\Delta E_{\text{sig}}$ cut efficiency after applying a double-Gaussian smearing to the signal MC sample. The observed efficiency variation, $0.1\%$ is taken as the systematic uncertainty.

\item Quantum correlation correction:
The uncertainties associated with the quantum-correlation parameters $r$, $R$, $\delta$ and $F_+$ are propagated according to the results of Refs.~\cite{QCC1,QCC2,BESIII:2020rxv}, resulting in a relative uncertainty of 0.1\%.

\end{itemize}

After correcting the differences in $K^{\pm}$ tracking and $\pi^0$ reconstruction efficiencies between data and MC simulation, the BF of $D^0 \to K^+ K^-\pi^0\pi^0$ is determined to be $\mathcal{B}(D^0 \to K^+ K^-\pi^0\pi^0)$ = \BF. 

\begin{table}[htbp]
  \begin{center}
    \begin{tabular}{|lc|}
      \hline
      Source   &Uncertainty (\%)\\
      \hline
      ST $\bar D^{0}$ candidates  & 0.3 \\
      Tracking                    & 0.2 \\
      PID                         &0.2  \\
      $\pi^0$ reconstruction      & 0.4 \\
      MC sample size              & 0.1 \\
      Amplitude  model            & 0.3 \\
      Quoted ${\mathcal B}(\pi^0\to\gamma\gamma)$    & 0.1\\
      2D fit   		            & 0.3 \\
          $\theta_{D^0\bar{D}^0}$ &0.5  \\
      $\Delta E_{\rm sig}$ requirement          & 0.1\\
      Quantum correlation correction	          & 0.1\\
      \hline
      Total                       & 0.9\\
      \hline
    \end{tabular}
  \end{center}
  \caption{Relative systematic uncertainties in the BF measurement.}
   \label{BF-Sys}
\end{table}

\section{Summary}

The first amplitude analysis of the decay $D^0 \to K^+K^-\pi^0\pi^0$ has been performed using 20.3 $\rm{fb}^{-1}$ of $e^+e^-$ collision data collected with the BESIII detector at the center-of-mass energy of 3.773 GeV. The BF of $D^0 \to K^+K^-\pi^0\pi^0$ is determined to be \BF. The result is consistent with the previous BESIII measurement of $(0.69 \pm 0.07_{\rm stat.} \pm 0.04_{\rm syst.})\times10^{-3}$~\cite{BESIII:2020rxv} within 1.0$\sigma$, with an improved precision by a factor of 2.5. Combining the FFs listed in Table~\ref{tab:signi}, the BFs for the intermediate processes are calculated from $\mathcal{B}_i = {\rm FF}_i \times \mathcal{B}(D^0\to K^+K^-\pi^0\pi^0)$, and summarized in Table~\ref{sum:FF_absBF}. 

According to the amplitude analysis, the dominant intermediate process is $D^0 \to K^{*}(892)^+K^{*}(892)^-\to K^+K^-\pi^0\pi^0$, with a BF of $(2.90\pm0.34_{\rm{stat.}}\pm0.12_{\rm{syst.}})\times 10^{-4}$. The absolute BF of $D^0 \to K^{*}(892)^+K^{*}(892)^-$ is determined to be $(2.61\pm0.31_{\rm{stat.}}\pm0.15_{\rm{syst.}})\times 10^{-3}$. 
Compared with the predictions listed in Table~\ref{tab:compare}, our result is significantly lower than those from Refs.~\cite{Cao:2023csx,Cheng:2010rv}. However, it aligns well with the prediction based on the flavor SU(3) symmetry model~\cite{Kamal:1990ky}.
The $D^0\to K^{*}(892)^+ K^{*}(892)^-$ decay is found to be $S$-wave dominant. The longitudinal polarization fraction $F_L$ is measured to be $F_L(D^0\to K^{*}(892)^+ K^{*}(892)^-)=0.468\pm0.046_{\rm{stat.}}\pm0.011_{\rm{syst.}}$, which is higher than the prediction in Ref.~\cite{Cao:2023csx} by more than three standard deviations. 
This polarization information in SCS $D$ decays provides insight into the understanding of hadronic decay dynamics and FSIs.

\begin{table}[htbp]
  \centering
    \begin{tabular}{l r@{$\pm$}c@{$\pm$}l }
    \hline
    \hline
    Intermediate process  &\multicolumn{3}{c}{BF~($\times 10^{-4}$)}\\
    \hline	
    $D^0\to K^{*}(892)^+ K^{*}(892)^-,K^{*}(892)^+\to K^+\pi^0, K^{*}(892)^-\to K^-\pi^0$
    &2.90  &0.34 &0.12 \\
    $D^0\to K_1(1400)^+ K^{-}, K_1(1400)^+\to K^{*}(892)^+ \pi^0, K^{*}(892)^+\to K^+\pi^0$
    &0.59  &0.18 &0.11\\
    $D^0\to f_1(1420)\pi^{0}, f_1(1420)\to K^{*}(892) K, K^{*}(892)\to K\pi^0$
    &0.58 &0.16 &0.08 \\
    $D^0\to \eta(1405)\pi^{0}, \eta(1405)\to a_0(980) \pi^0, a_0(980)\to K^+K^-$
    &1.37 &0.20 &0.14 \\
    $D^0\to \eta(1475)\pi^{0}, \eta(1475)\to K^{*}(892) K, K^{*}(892)\to K\pi^0$
    &2.42 &0.44 &0.18 \\
    \hline
    \hline
  \end{tabular}
   \caption{The BFs of various intermediate processes in $D^0 \to K^-K^+\pi^0\pi^0$. The first and second uncertainties are statistical and systematic, respectively. }
  \label{sum:FF_absBF}
\end{table}

\acknowledgments

The BESIII Collaboration thanks the staff of BEPCII (https://cstr.cn/31109.02.BEPC) and the IHEP computing center for their strong support. This work is supported in part by National Key R\&D Program of China under Contracts Nos. 2023YFA1606000, 2023YFA1606704; National Natural Science Foundation of China (NSFC) under Contracts Nos. 11635010, 11935015, 11935016, 11935018, 12025502, 12035009, 12035013, 12061131003, 12192260, 12192261, 12192262, 12192263, 12192264, 12192265, 12221005, 12225509, 12235017, 12342502, 12361141819; the Excellent Youth Foundation of Henan Scientific Commitee under Contract No.~242300421044; the Chinese Academy of Sciences (CAS) Large-Scale Scientific Facility Program; the Strategic Priority Research Program of Chinese Academy of Sciences under Contract No. XDA0480600; Joint Large-Scale Scientific Facility Fund of the NSFC and the Chinese Academy of Sciences under Contract No.~U2032104; CAS under Contract No. YSBR-101; 100 Talents Program of CAS; The Institute of Nuclear and Particle Physics (INPAC) and Shanghai Key Laboratory for Particle Physics and Cosmology; ERC under Contract No. 758462; German Research Foundation DFG under Contract No. FOR5327; Istituto Nazionale di Fisica Nucleare, Italy; Knut and Alice Wallenberg Foundation under Contracts Nos. 2021.0174, 2021.0299, 2023.0315; Ministry of Development of Turkey under Contract No. DPT2006K-120470; National Research Foundation of Korea under Contract No. NRF-2022R1A2C1092335; National Science and Technology fund of Mongolia; Polish National Science Centre under Contract No. 2024/53/B/ST2/00975; STFC (United Kingdom); Swedish Research Council under Contract No. 2019.04595; U. S. Department of Energy under Contract No. DE-FG02-05ER41374

\bibliographystyle{JHEP}
\bibliography{references}

@article{Bajc_1997,
   title={Nonleptonic two-body charmed meson decays in an effective model for their semileptonic decays},
   volume={56},
   ISSN={1089-4918},
   url={http://dx.doi.org/10.1103/PhysRevD.56.7207},
   DOI={10.1103/physrevd.56.7207},
   number={11},
   journal={Physical Review D},
   publisher={American Physical Society (APS)},
   author={Bajc, B. and Fajfer, S. and Oakes, R. J. and Prelovšek, S.},
   year={1997},
   month=dec, pages={7207–7215} }

@article{BESIII:2022udq,
    author = "Ablikim, Medina and others",
    collaboration = "BESIII",
    title = "{Partial wave analysis of the charmed baryon hadronic decay $ {\Lambda}_c^{+} ${\textrightarrow} {\ensuremath{\Lambda}}{\ensuremath{\pi}}$^{+}${\ensuremath{\pi}}$^{0}$}",
    eprint = "2209.08464",
    archivePrefix = "arXiv",
    primaryClass = "hep-ex",
    doi = "10.1007/JHEP12(2022)033",
    journal = "JHEP",
    volume = "12",
    pages = "033",
    year = "2022"
}

@article{BESIII:2020rxv,
    author = "Ablikim, Medina and others",
    collaboration = "BESIII",
    title = "{Measurements of the absolute branching fractions of $D^{0(+)}\to K\bar K\pi\pi$ decays}",
    eprint = "2007.10563",
    archivePrefix = "arXiv",
    primaryClass = "hep-ex",
    doi = "10.1103/PhysRevD.102.052006",
    journal = "Phys. Rev. D",
    volume = "102",
    number = "5",
    pages = "052006",
    year = "2020"
}

@article{Cheng:2010rv,
    author = "Cheng, Hai-Yang and Chiang, Cheng-Wei",
    title = "{Long-Distance Contributions to $D^0-\bar{D}^0$ Mixing Parameters}",
    eprint = "1005.1106",
    archivePrefix = "arXiv",
    primaryClass = "hep-ph",
    doi = "10.1103/PhysRevD.81.114020",
    journal = "Phys. Rev. D",
    volume = "81",
    pages = "114020",
    year = "2010"
}

@article{Cao:2023csx,
    author = "Cao, Ye and Cheng, Yin and Zhao, Qiang",
    title = "{Resolving the polarization puzzles in $D^0 \rightarrow VV$}",
    eprint = "2303.00535",
    archivePrefix = "arXiv",
    primaryClass = "hep-ph",
    doi = "10.1103/PhysRevD.109.073002",
    journal = "Phys. Rev. D",
    volume = "109",
    number = "7",
    pages = "073002",
    year = "2024"
}

@article{Kamal:1990ky,
    author = "Kamal, A. N. and Verma, R. C. and Sinha, N.",
    title = "{$D/D_s^{\pm} \to VV$ decays in two models: An SU(3) symmetry model and a factorization model with final state interactions}",
    reportNumber = "ALBERTA-THY-18-90",
    doi = "10.1103/PhysRevD.43.843",
    journal = "Phys. Rev. D",
    volume = "43",
    pages = "843--854",
    year = "1991"
}

@article{Cheng:2011pb,
    author = "Cheng, Hai-Yang",
    title = "{Revisiting Axial-Vector Meson Mixing}",
    eprint = "1110.2249",
    archivePrefix = "arXiv",
    primaryClass = "hep-ph",
    doi = "10.1016/j.physletb.2011.12.013",
    journal = "Phys. Lett. B",
    volume = "707",
    pages = "116--120",
    year = "2012"
}

@article{Cheng:2013cwa,
    author = "Cheng, Hai-Yang",
    editor = "Iijima, Toru",
    title = "{Mixing angle of $K_1$ axial vector mesons}",
    eprint = "1311.2370",
    archivePrefix = "arXiv",
    primaryClass = "hep-ph",
    doi = "10.22323/1.205.0090",
    journal = "PoS",
    volume = "Hadron2013",
    pages = "090",
    year = "2013"
}

@article{Dunietz:1990cj,
    author = "Dunietz, Isard and Quinn, Helen R. and Snyder, A. and Toki, W. and Lipkin, Harry J.",
    title = "{How to extract CP violating asymmetries from angular correlations}",
    reportNumber = "SLAC-PUB-5270",
    doi = "10.1103/PhysRevD.43.2193",
    journal = "Phys. Rev. D",
    volume = "43",
    pages = "2193--2208",
    year = "1991"
}

@article{Valencia:1988it,
    author = "Valencia, German",
    title = "{Angular Correlations in the Decay $B \to V V$ and {CP} Violation}",
    reportNumber = "BNL-42229",
    doi = "10.1103/PhysRevD.39.3339",
    journal = "Phys. Rev. D",
    volume = "39",
    pages = "3339",
    year = "1989"
}

@article{PDG,
  title = {Review of Particle Physics},
  author = {Navas, S. and Amsler, C. and Gutsche, T. and Hanhart, C. and Hern\'andez-Rey, J. J. and Louren\ifmmode \mbox{\c{c}}\else \c{c}\fi{}o, C. and Masoni, A. and Mikhasenko, M. and Mitchell, R. E. and Patrignani, C. and Schwanda, C. and Spanier, S. and Venanzoni, G. and Yuan, C. Z. and Agashe, K. and Aielli, G. and Allanach, B. C. and Alvarez-Mu\~niz, J. and Antonelli, M. and Aschenauer, E. C. and Asner, D. M. and Assamagan, K. and Baer, H. and Banerjee, Sw. and Barnett, R. M. and Baudis, L. and Bauer, C. W. and Beatty, J. J. and Beringer, J. and Bettini, A. and Biebel, O. and Black, K. M. and Blucher, E. and Bonventre, R. and Briere, R. A. and Buckley, A. and Burkert, V. D. and Bychkov, M. A. and Cahn, R. N. and Cao, Z. and Carena, M. and Casarosa, G. and Ceccucci, A. and Cerri, A. and Chivukula, R. S. and Cowan, G. and Cranmer, K. and Crede, V. and Cremonesi, O. and D'Ambrosio, G. and Damour, T. and de Florian, D. and de Gouv\^ea, A. and DeGrand, T. and Demers, S. and Demiragli, Z. and Dobrescu, B. A. and D'Onofrio, M. and Doser, M. and Dreiner, H. K. and Eerola, P. and Egede, U. and Eidelman, S. and El-Khadra, A. X. and Ellis, J. and Eno, S. C. and Erler, J. and Ezhela, V. V. and Fava, A. and Fetscher, W. and Fields, B. D. and Freitas, A. and Gallagher, H. and Gershon, T. and Gershtein, Y. and Gherghetta, T. and Gonzalez-Garcia, M. C. and Goodman, M. and Grab, C. and Gritsan, A. V. and Grojean, C. and Groom, D. E. and Gr\"unewald, M. and Gurtu, A. and Haber, H. E. and Hamel, M. and Hashimoto, S. and Hayato, Y. and Hebecker, A. and Heinemeyer, S. and Hikasa, K. and Hisano, J. and H\"ocker, A. and Holder, J. and Hsu, L. and Huston, J. and Hyodo, T. and Ianni, Al. and Kado, M. and Karliner, M. and Katz, U. F. and Kenzie, M. and Khoze, V. A. and Klein, S. R. and Krauss, F. and Kreps, M. and Kri\ifmmode \check{z}\else \v{z}\fi{}an, P. and Krusche, B. and Kwon, Y. and Lahav, O. and Lellouch, L. P. and Lesgourgues, J. and Liddle, A. R. and Ligeti, Z. and Lin, C.-J. and Lippmann, C. and Liss, T. M. and Lister, A. and Littenberg, L. and Lugovsky, K. S. and Lugovsky, S. B. and Lusiani, A. and Makida, Y. and Maltoni, F. and Manohar, A. V. and Marciano, W. J. and Matthews, J. and Mei\ss{}ner, U.-G. and Melzer-Pellmann, I.-A. and Mertsch, P. and Miller, D. J. and Milstead, D. and M\"onig, K. and Molaro, P. and Moortgat, F. and Moskovic, M. and Nagata, N. and Nakamura, K. and Narain, M. and Nason, P. and Nelles, A. and Neubert, M. and Nir, Y. and O'Connell, H. B. and O'Hare, C. A. J. and Olive, K. A. and Peacock, J. A. and Pianori, E. and Pich, A. and Piepke, A. and Pietropaolo, F. and Pomarol, A. and Pordes, S. and Profumo, S. and Quadt, A. and Rabbertz, K. and Rademacker, J. and Raffelt, G. and Ramsey-Musolf, M. and Richardson, P. and Ringwald, A. and Robinson, D. J. and Roesler, S. and Rolli, S. and Romaniouk, A. and Rosenberg, L. J and Rosner, J. L. and Rybka, G. and Ryskin, M. G. and Ryutin, R. A. and Safdi, B. and Sakai, Y. and Sarkar, S. and Sauli, F. and Schneider, O. and Sch\"onert, S. and Scholberg, K. and Schwartz, A. J. and Schwiening, J. and Scott, D. and Sefkow, F. and Seljak, U. and Sharma, V. and Sharpe, S. R. and Shiltsev, V. and Signorelli, G. and Silari, M. and Simon, F. and Sj\"ostrand, T. and Skands, P. and Skwarnicki, T. and Smoot, G. F. and Soffer, A. and Sozzi, M. S. and Spiering, C. and Stahl, A. and Sumino, Y. and Takahashi, F. and Tanabashi, M. and Tanaka, J. and Ta\ifmmode \check{s}\else \v{s}\fi{}evsk\'y, M. and Terao, K. and Terashi, K. and Terning, J. and Thoma, U. and Thorne, R. S. and Tiator, L. and Titov, M. and Tovey, D. R. and Trabelsi, K. and Urquijo, P. and Valencia, G. and Van de Water, R. and Varelas, N. and Verde, L. and Vivarelli, I. and Vogel, P. and Vogelsang, W. and Vorobyev, V. and Wakely, S. P. and Walkowiak, W. and Walter, C. W. and Wands, D. and Weinberg, D. H. and Weinberg, E. J. and Wermes, N. and White, M. and Wiencke, L. R. and Willocq, S. and Woody, C. L. and Workman, R. L. and Yao, W.-M. and Yokoyama, M. and Yoshida, R. and Zanderighi, G. and Zeller, G. P. and Zhu, R.-Y. and Zhu, S.-L. and Zimmermann, F. and Zyla, P. A. and Anderson, J. and Kramer, M. and Schaffner, P. and Zheng, W.},
  collaboration = {Particle Data Group},
  journal = {Phys. Rev. D},
  volume = {110},
  issue = {3},
  pages = {030001},
  numpages = {5},
  year = {2024},
  month = {Aug},
  publisher = {American Physical Society},
  doi = {10.1103/PhysRevD.110.030001},
  url = {https://link.aps.org/doi/10.1103/PhysRevD.110.030001}
}

@article{Ablikim:2009aa,
title = {Design and construction of the {BESIII} detector},
journal = "Nucl. Instrum. Meth. A",
volume = {614},
number = {3},
pages = {345-399},
year = {2010},
issn = {0168-9002},
doi = {https://doi.org/10.1016/j.nima.2009.12.050},
url = {https://www.sciencedirect.com/science/article/pii/S0168900209023870},
author = {M. Ablikim and Z.H. An and J.Z. Bai and Niklaus Berger and J.M. Bian and X. Cai and G.F. Cao and X.X. Cao and J.F. Chang and C. Chen and G. Chen and H.C. Chen and H.X. Chen and J. Chen and J.C. Chen and L.P. Chen and P. Chen and X.H. Chen and Y.B. Chen and M.L. Chen and Y.P. Chu and X.Z. Cui and H.L. Dai and Z.Y. Deng and M.Y. Dong and S.X. Du and Z.Z. Du and J. Fang and C.D. Fu and C.S. Gao and M.Y. Gong and W.X. Gong and S.D. Gu and B.J. Guan and J. Guan and Y.N. Guo and J.F. Han and K.L. He and M. He and X. He and Y.K. Heng and Z.L. Hou and H.M. Hu and T. Hu and B. Huang and J. Huang and S.K. Huang and Y.P. Huang and Q. Ji and X.B. Ji and X.L. Ji and L.K. Jia and L.L. Jiang and X.S. Jiang and D.P. Jin and S. Jin and Y. Jin and Y.F. Lai and G.K. Lei and F. Li and G. Li and H.B. Li and H.S. Li and J. Li and J. Li and J.C. Li and Q.J. Li and L. Li and L. Li and R.B. Li and R.Y. Li and W.D. Li and W.G. Li and X.N. Li and X.P. Li and X.R. Li and Y.R. Li and W. Li and D.X. Lin and B.J. Liu and C.X. Liu and F. Liu and G.M. Liu and H. Liu and H.M. Liu and H.W. Liu and J.B. Liu and L.F. Liu and Q. Liu and Q.G. Liu and S.D. Liu and W.J. Liu and X. Liu and X.Z. Liu and Y. Liu and Y.J. Liu and Z.A. Liu and Z.Q. Liu and Z.X. Liu and J.G. Lu and T. Lu and Y.P. Lu and X.L. Luo and H.L. Ma and Q.M. Ma and X. Ma and X.Y. Ma and Z.P. Mao and J. Min and X.H. Mo and J. Nie and Z.D. Nie and R.G. Ping and S. Qian and Q. Qiao and G. Qin and Z.H. Qin and J.F. Qiu and R.G. Liu and Z.Y. Ren and G. Rong and L. Shang and D.L. Shen and X.Y. Shen and H.Y. Sheng and Y.F. Shi and L.W. Song and W.Y. Song and D.H. Sun and G.X. Sun and H.S. Sun and L.J. Sun and S.S. Sun and X.D. Sun and Y.Z. Sun and Z.J. Sun and J.P. Tan and S.Q. Tang and X. Tang and N. Tao and H.L. Tian and Y.R. Tian and X. Wan and D.Y. Wang and J.K. Wang and J.Z. Wang and K. Wang and K.X. Wang and L. Wang and L. Wang and L.J. Wang and L.S. Wang and M. Wang and N. Wang and P. Wang and P.L. Wang and Q. Wang and Y.F. Wang and Z. Wang and Z. Wang and Z.G. Wang and Z.Y. Wang and C.L. Wei and S.J. Wei and S.P. Wen and J.J. Wu and L.H. Wu and N. Wu and Y.H. Wu and Y.M. Wu and Z. Wu and M.H. Xu and X.M. Xia and H.S. Xiang and G. Xie and X.X. Xie and Y.G. Xie and G.F. Xu and H. Xu and Q.J. Xu and J.D. Xue and L. Xue and L. Yan and G.A. Yang and H. Yang and H.X. Yang and S.M. Yang and M. Ye and B.X. Yu and C. Yuan and C.Z. Yuan and Y. Yuan and S.L. Zang and B.X. Zhang and B.Y. Zhang and C.C. Zhang and C.C. Zhang and D.H. Zhang and H.Y. Zhang and J. Zhang and J.W. Zhang and J.Y. Zhang and L.S. Zhang and M. Zhang and Q.X. Zhang and W. Zhang and X.M. Zhang and Y. Zhang and Y.H. Zhang and Y.Y. Zhang and Z.X. Zhang and S.H. Zhang and D.X. Zhao and D.X. Zhao and H.S. Zhao and J.B. Zhao and J.W. Zhao and J.Z. Zhao and L. Zhao and P.P. Zhao and Y.B. Zhao and Y.D. Zhao and B. Zheng and J.P. Zheng and L.S. Zheng and Z.P. Zheng and B.Q. Zhou and G.M. Zhou and J. Zhou and L. Zhou and Z.L. Zhou and H.T. Zhu and K. Zhu and K.J. Zhu and Q.M. Zhu and X.W. Zhu and Y.S. Zhu and Z.A. Zhu and B.A. Zhuang and J.H. Zou and X. Zou and J.X. Zuo and L.L. Wang and M.H. Ye and Y.H. Zheng and C.F. Qiao and X.R. Lu and H.B. Liu and J.F. Hu and Y.T. Gu and X.D. Ruan and G.M. Huang and Y. Zeng and Y.H. Yan and G. Chelkov and I. Boyko and D. Dedovich and I. Denysenko and S. Grishin and A. Zhemchugov and Zhenjun Xiao and Jialun Ping and Libo {Guo Chenglin Luo} and Shenjian Chen and Ming Qi and Xiaowei Hu and Lei Zhang and Xueqian Li and Chunxu Yu and Yubin Liu and Ye Xu and Minggang Zhao and Aiqiang Guo and Yuping Guo and Zhenya He and Y.J. Mao and Z.Y. You and Y.T. Liang and X.Y. Zhang and X.T. Huang and J.B. Jiao and X.L. Li and M.Y. Duan and F.H. Liu and Q.W. Lu and F.P. Ning and X.D. Wang and Yongfei Liang and Changjian Tang and Yiyun Zhang and Y.N. Gao and H. Gong and B.B. Shao and Y.R. Tian and S.M. Yang and F.A. Harris and J.W. Kennedy and Q. Liu and X. Nguyen and S.L. Olsen and M. Rosen and C.P. Shen and G.S. Varner and X. Yu and Y. Zhou and H. Liang and Y. Chen and J. Xue and Q. Liu and B. Liu and Z. Cheng and L. Zhou and H. Yang and H.F. Chen and Cheng Li and M. Shao and Y.J. Sun and J. Yan and Z.B. Tang and X. Li and L. Zhao and L. Jiang and Z.P. Zhang and J. Wu and Z.Z. Xu and Q. Shan and Z. Xue and X.L. Wang and Q. An and S.B. Liu and J.H. Guo and L. Zhao and C.Q. Feng and X.Z. Liu and H. Li and W. Zheng and H. Yan and Z. Cao and X.H. Liu and Sachio Komamiya and Tomoyuki Sanuki and Taiki Yamamura and Tianchi Zhao and Mingxing Luo}
}

@article{Ablikim:2019hff,

    collaboration = "BESIII",
    author = "Ablikim, M. and others ",
title = "{Future physics programme of BESIII}",
    archivePrefix = "arXiv",
    primaryClass = "hep-ex",
    reportNumber = "HEP-Physics-Report-BESIII-2019-12-13",
    doi = "10.1088/1674-1137/44/4/040001",
    journal = "Chin. Phys. C",
    volume = "44",
    number = "4",
    pages = "040001",
    year = "2020"
}

@InProceedings{Yu:IPAC2016-TUYA01,
  author       = {C.H. Yu and others},
  title        = {{BEPCII} {P}erformance and {B}eam {D}ynamics {S}tudies on {L}uminosity},
  booktitle    = {Proc. of International Particle Accelerator Conference (IPAC'16),
                  Busan, Korea, May 8-13, 2016},
  pages        = {1014--1018},
  paper        = {TUYA01},
  language     = {english},
  venue        = {Busan, Korea},
  series       = {International Particle Accelerator Conference},
  number       = {7},
  publisher    = {JACoW},
  address      = {Geneva, Switzerland},
  month        = {June},
  year         = {2016},
  isbn         = {978-3-95450-147-2},
  doi          = {doi:10.18429/JACoW-IPAC2016-TUYA01},
  url          = {http://jacow.org/ipac2016/papers/tuya01.pdf},
  note         = {doi:10.18429/JACoW-IPAC2016-TUYA01},
}

@article{etof1,
    author = "Li, Xin and Sun, Yongjie and Li, Cheng and Liu, Zhen and Heng, Yuekun and Shao, Ming and Wang, Xiaozhuang and Wu, Zhi and Cao, Ping and Chen, Mingming and Dai, Hongliang and Liu, Shubing and Luo, Xiaolan and Jiang, Xiaoshan and Sun, Shengsen and Tang, Zebo and Sun, Weijia and Wang, Siyu and Xu, Meihang and Yang, Rongxing and Zhu, Kejun",
    title = "{Study of MRPC technology for BESIII endcap-TOF upgrade}",
    doi = "10.1007/s41605-017-0014-2",
    journal = "Radiat. Detect. Technol. Methods",
    volume = "1",
    pages = "13",
    year = "2017"
}

@article{etof2,
    author = "Guo, Ying-Xiao and Sun, Sheng-Sen and An, Fen-Fen and Yang, Rong-Xing and Zhou, Ming and Wu, Zhi and Dai, Hong-Liang and Heng, Yue-Kun and Li, Cheng and Deng, Zi-Yan and Liu, Huai-Min and Li, Wei-Guo",
    title = "{The study of time calibration for upgraded end cap TOF of BESIII}",
    doi = "10.1007/s41605-017-0012-4",
    journal = "Radiat. Detect. Technol. Methods",
    volume = "1",
    pages = "15",
    year = "2017"
}

@article{etof3,
title = {Design and construction of the new BESIII endcap Time-of-Flight system with MRPC Technology},
journal = {Nucl. Instrum. Meth. A},
volume = {953},
pages = {163053},
year = {2020},
issn = {0168-9002},
doi = {https://doi.org/10.1016/j.nima.2019.163053},
url = {https://www.sciencedirect.com/science/article/pii/S0168900219314068},
author = {P. Cao and H.F. Chen and M.M. Chen and H.L. Dai and Y.K. Heng and X.L. Ji and X.S. Jiang and C. Li and X. Li and S.B. Liu and Z. Liu and X.L. Luo and X.Y. Ma and M. Shao and S.S. Sun and Y.J. Sun and Z.B. Tang and X.Z. Wang and Z. Wu and M.H. Xu and R.X. Yang and M. Ye and J. Zhang and Y.H. Zhang and J.Z. Zhao}
}

@article{GEANT4,
    author = "Agostinelli, S. and others",
    collaboration = "GEANT4",
    title = "{GEANT4 -- a simulation toolkit}",
    reportNumber = "SLAC-PUB-9350, FERMILAB-PUB-03-339, CERN-IT-2002-003",
    doi = "10.1016/S0168-9002(03)01368-8",
    journal = "Nucl. Instrum. Meth. A",
    volume = "506",
    pages = "250--303",
    year = "2003"
}

@article{KKMC1,
  title = {Coherent exclusive exponentiation for precision {M}onte {C}arlo calculations},
  author = {Jadach, S. and Ward, B. F. L. and Wa\ifmmode \mbox{\c{}}\else \c{}\fi{}s, Z.},
  journal = {Phys. Rev. D},
  volume = {63},
  issue = {11},
  pages = {113009},
  numpages = {65},
  year = {2001},
  month = {May},
  publisher = {American Physical Society},
  doi = {10.1103/PhysRevD.63.113009},
  url = {https://link.aps.org/doi/10.1103/PhysRevD.63.113009}
}

@article{KKMC2,
title = {The precision Monte Carlo event generator {KK} for two-fermion final states in $e^+e^-$ collisions},
journal = {Computer Physics Communications},
volume = {130},
number = {3},
pages = {260-325},
year = {2000},
issn = {0010-4655},
doi = {https://doi.org/10.1016/S0010-4655(00)00048-5},
url = {https://www.sciencedirect.com/science/article/pii/S0010465500000485},
author = {Jadach, S. and Ward, B. F. L. and Wa\ifmmode \mbox{\c{}}\else \c{}\fi{}s, Z.}
}

@article{EVTGEN1,
    author = "Lange, D.J.",
    editor = "Erhan, S. and Schlein, P. and Rozen, Y.",
    title = "{The EvtGen particle decay simulation package}",
    doi = "10.1016/S0168-9002(01)00089-4",
    journal = "Nucl. Instrum. Meth. A",
    volume = "462",
    pages = "152--155",
    year = "2001"
}

@article{EVTGEN2,
        doi = {10.1088/1674-1137/32/8/001},
        url = {https://doi.org/10.1088%2F1674-1137%2F32%2F8%2F001},
        year = 2008,
        month = {aug},
        publisher = {{IOP} Publishing},
        volume = {32},
        number = {8},
        pages = {599--602},
        author = {Ping, Rong-Gang},
        title = {Event generators at {BESIII}},
        journal = {Chin. Phys. C}
}

@article{LUNDCHARM1,
  title = "{Event generator for $J/\ensuremath{\psi}$ and $\ensuremath{\psi}(2S)$ decay}",
  author = {Chen, J. C. and Huang, G. S. and Qi, X. R. and Zhang, D. H. and Zhu, Y. S.},
  journal = {Phys. Rev. D},
  volume = {62},
  issue = {3},
  pages = {034003},
  numpages = {8},
  year = {2000},
  month = {Jun},
  publisher = {American Physical Society},
  doi = {10.1103/PhysRevD.62.034003},
  url = {https://link.aps.org/doi/10.1103/PhysRevD.62.034003}
}

@article{LUNDCHARM2,
author = {Yang, Rui-Ling and Ping, Rong-Gang and Chen, Hong},
title = "{Tuning and validation of the Lundcharm model with $J/\psi$ decays}",
publisher = {Chin. Phys. Lett.},
year = {2014},
journal = {Chin. Phys. Lett.},
volume = {31},
number = {06},
eid = {061301},
numpages = {0},
pages = {061301},
url = {http://cpl.iphy.ac.cn/EN/abstract/article_60019.shtml},
doi = {10.1088/0256-307X/31/6/061301}
}

@article{PHOTOS,
title = "{QED bremsstrahlung in semileptonic B and leptonic $\tau$ decays}",
journal = "Phys. Lett. B",
volume = "303",
number = "1",
pages = "163 - 169",
year = "1993",
issn = "0370-2693",
doi = "https://doi.org/10.1016/0370-2693(93)90062-M",
url = "http://www.sciencedirect.com/science/article/pii/037026939390062M",
author = "E. Richter-Was"
}

@article{covariant-tensors,
	author = {{B.S. Zou} and {D.V. Bugg}},
	title = {Covariant tensor formalism for partial-wave analyses of $\psi$ decay to mesons},
	DOI= "10.1140/epja/i2002-10135-4",
	url= "https://doi.org/10.1140/epja/i2002-10135-4",
	journal = {Eur. Phys. J. A},
	year = 2003,
	volume = 16,
	number = 4,
	pages = "537-547",
	month = "",
}

@article{EventFilter,
    author = "Zhang, Jia-Wei and others",
    title = "{Suppression of top-up injection backgrounds with offline event filter in the BESIII experiment}",
    doi = "10.1007/s41605-022-00331-7",
    journal = "Radiat. Detect. Technol. Methods",
    volume = "6",
    number = "3",
    pages = "289--293",
    year = "2022"
}

@article{KPsnew1,
    author = "Aston, D. and others",
    title = "{A Study of $K^- \pi^+$ Scattering in the Reaction $K^- p \to K^- \pi^+ n$  at $11~{\rm GeV}/c$}",
    reportNumber = "SLAC-PUB-4260, DPNU-87-25",
    doi = "10.1016/0550-3213(88)90028-4",
    journal = "Nucl. Phys. B",
    volume = "296",
    pages = "493--526",
    year = "1988"
}

@article{Kpsnew2,
    author = "Adachi, I. and others",
    collaboration = "BaBar, Belle",
    title = "{Measurement of $\cos{2\beta}$ in $B^{0} \to D^{(*)} h^{0}$ with $D \to K_{S}^{0} \pi^{+} \pi^{-}$ decays by a combined time-dependent Dalitz plot analysis of BaBar and Belle data}",
    archivePrefix = "arXiv",
    primaryClass = "hep-ex",
    reportNumber = "BABAR-PUB-18-002",
    doi = "10.1103/PhysRevD.98.112012",
    journal = "Phys. Rev. D",
    volume = "98",
    number = "11",
    pages = "112012",
    year = "2018"
}

@article{Luminosity,
    author = "Ablikim, Medina and others",
    collaboration = "BESIII",
    title = "{Measurement of $e^+ e^{-} \rightarrow D\bar D$ cross sections at the $\psi(3770)$ resonance}",
    archivePrefix = "arXiv",
    primaryClass = "hep-ex",
    doi = "10.1088/1674-1137/42/8/083001",
    journal = "Chin. Phys. C",
    volume = "42",
    number = "8",
    pages = "083001",
    year = "2018"
}

@article{QCC1,
    author = "Aaij, Roel and others", 
    collaboration = "LHCb",
    title = "{Simultaneous determination of the CKM angle $\gamma$ and parameters related to mixing and \emph{CP} violation in the charm sector}",
    journal = "\href{https://cds.cern.ch/record/2905625}{LHCb-CONF-2024-004}",
    year = "2024"
}

@article{QCC2,
    author = "Ablikim, Medina and others",
    collaboration = "BESIII",
    title = "{Measurement of the $D \to K^-\pi^+\pi^+\pi^-$ and $D \to K^-\pi^+\pi^0$ coherence factors and average strong-phase differences in quantum-correlated ${D\bar{D}}$ decays}",
    archivePrefix = "arXiv",
    primaryClass = "hep-ex",
    doi = "10.1007/JHEP05(2021)164",
    journal = "JHEP",
    volume = "05",
    pages = "164",
    year = "2021"
}

@article{ARGUS,
    author = "Albrecht, H. and others",
    collaboration = "ARGUS",
    title = "{Search for Hadronic $b \to u$ Decays}",
    reportNumber = "DESY-90-008",
    doi = "10.1016/0370-2693(90)91293-K",
    journal = "Phys. Lett. B",
    volume = "241",
    pages = "278--282",
    year = "1990"
}

@article{lum1,
    author = "Ablikim, M. and others",
    collaboration = "BESIII",
    title = "{Measurement of the integrated luminosities of the data taken by BESIII at $\sqrt{s}=$3.650 and 3.773 GeV}",
    archivePrefix = "arXiv",
    primaryClass = "hep-ex",
    doi = "10.1088/1674-1137/37/12/123001",
    journal = "Chin. Phys. C",
    volume = "37",
    pages = "123001",
    year = "2013"
}

@article{lum2,

    collaboration = "BESIII",
    author = "Ablikim, Medina and others",
    title = "{Measurement of integrated luminosity of data collected at 3.773 GeV by BESIII from 2021 to 2024}",
    archivePrefix = "arXiv",
    primaryClass = "hep-ex",
    doi = "10.1088/1674-1137/ad70a0",
    journal = "Chin. Phys. C",
    volume = "48",
    number = "12",
    pages = "123001",
    year = "2024"
}

@article{EcmsMea,
title={Online monitoring of the center-of-mass energy from real data at {BESIII}},
author={Lu, Jiada and Xiao, Yanjia and Ji, Xiaobin},
journal={Radiation Detection Technology and Methods},
volume={4},
number={3},
pages={337-344},
year={2020},
doi={10.1007/s41605-020-00188-8},
issn={2509-9949},
url={https://doi.org/10.1007/s41605-020-00188-8}
}

@article{TDA_QCD,
    author = "Cheng, Hai-Yang and Chiang, Cheng-Wei",
    title = "{Revisiting CP violation in $D\to P\!P$ and $V\!P$ decays}",
    archivePrefix = "arXiv",
    primaryClass = "hep-ph",
    doi = "10.1103/PhysRevD.100.093002",
    journal = "Phys. Rev. D",
    volume = "100",
    number = "9",
    pages = "093002",
    year = "2019"
}

@article{Cheng,
    author = "Cheng, Hai-Yang and Chiang, Cheng-Wei",
    title = "{Two-body hadronic charmed meson decays}",
    archivePrefix = "arXiv",
    primaryClass = "hep-ph",
    doi = "10.1103/PhysRevD.81.074021",
    journal = "Phys. Rev. D",
    volume = "81",
    pages = "074021",
    year = "2010"
}

@article{XG1,
    author = "Rogozhnikov, A.",
    editor = "Salinas, Luis and Torres, Claudio",
    title = "{Reweighting with Boosted Decision Trees}",
    archivePrefix = "arXiv",
    primaryClass = "physics.data-an",
    doi = "10.1088/1742-6596/762/1/012036",
    journal = "J. Phys. Conf. Ser.",
    volume = "762",
    number = "1",
    pages = "012036",
    year = "2016"
}

@article{XG2,
    author = "Liu, Beijiang and Xiong, Xian and Hou, Guoyi and Song, Shiming and Shen, Lin",
    editor = "Forti, A. and Betev, L. and Litmaath, M. and Smirnova, O. and Hristov, P.",
    title = "{Applications of Machine Learning at BESIII}",
    doi = "10.1051/epjconf/201921406033",
    journal = "EPJ Web Conf.",
    volume = "214",
    pages = "06033",
    year = "2019"
}

@article{DTmethod,
    author = "Baltrusaitis, R. M. and others",
    collaboration = "MARK-III",
    title = "{Direct Measurements of Charmed d Meson Hadronic Branching Fractions}",
    reportNumber = "SLAC-PUB-3861",
    doi = "10.1103/PhysRevLett.56.2140",
    journal = "Phys. Rev. Lett.",
    volume = "56",
    pages = "2140",
    year = "1986"
}

@book{Blatte,
  author    = {Blatt, John M. and Weisskopf, Victor F.},
  title     = {Theoretical Nuclear Physics},
  publisher = {John Wiley \& Sons},
  address   = {New York},
  year      = {1973},
  pages     = {864},
  isbn      = {978-0-471-09281-0}
}

@article{QCBF,
    author = "Ablikim, Medina and others",
    collaboration = "BESIII",
    title = "{Amplitude analysis of the decays $D^0 \to \pi^+ \pi^- \pi^+ \pi^-$ and $D^0 \to \pi^+ \pi^- \pi^0 \pi^0$}",
    archivePrefix = "arXiv",
    primaryClass = "hep-ex",
    doi = "10.1088/1674-1137/ad3d4d",
    journal = "Chin. Phys. C",
    volume = "48",
    number = "8",
    pages = "083001",
    year = "2024"
}

\clearpage
\appendix
\section{\texorpdfstring{$M_{\mathrm{BC}}^{\mathrm{sig}}$ versus $M_{\mathrm{BC}}^{\mathrm{tag}}$ two-dimensional fit}{MBC-sig versus MBC-tag two-dimensional fit}}

\label{2dfit}
The signal yields of DT candidates are determined by a 2D maximum likelihood binned fit to the distribution of $M_{\rm{BC}}^{\rm{sig}}$ versus $M_{\rm{BC}}^{\rm{tag}}$. Signal events with both tag and signal sides reconstructed correctly concentrate around $M_{\rm{BC}}^{\rm{sig}} = M_{\rm{BC}}^{\rm{tag}} = M_{D^0}$, where $M_{D^0}$ is the known $D^0$ mass~\cite{PDG}.  We define three kinds of background contributions. Candidates with correctly reconstructed $D^0$(or $\bar D^0$) and incorrectly reconstructed $\bar D^0$(or $D^0$) are BKGI, which appear around the bands $M_{\rm BC}^{\rm sig}$ or $M_{\rm BC}^{\rm tag} = M_{D^0}$. Other candidates appearing around the diagonal are mainly from the $D \bar D$ mispartition and the $e^+e^-\to q\bar{q}$ processes (BKGII). The rest of the flat background contributions mainly come from candidates reconstructed incorrectly on both sides (BKGIII). The PDFs for the different components used in the fit are given below:
\begin{itemize}
\item Signal: s($x, y$),
\item BKGI: $b_1(x,y)$,
\item BKGII:  $b_2(x,y)$,
\item BKGIII: argus($x; m_0, c, p$) $\cdot$ argus($y; m_0, c, p$).
\end{itemize}

The signal shape s($x, y$) is described by the 2D MC-simulated shape convolved with a 2D Gaussian. The parameters of the Gaussian function are obtained by a 1D fit on $M_{\rm{BC}}$ in the signal and the tag sides, respectively, and are fixed in the 2D fit. For BKGI and BKGII, $b_{1}(x,y)$ and $b_{2}(x,y)$ are both described by an MC-simulated shape. For BKGIII, it is constructed by an ARGUS function~\cite{ARGUS} in $M_{\rm{BC}}^{\rm{sig}}$ multiplied by an ARGUS function in $M_{\rm{BC}}^{\rm{tag}}$. In the fit, the parameters $m_0$ and $p$ for the ARGUS function~\cite{ARGUS} are fixed at 1.8865 GeV$/c^2$ and 0.5, respectively. 

\section{Longitudinal polarization fraction for $D \to VV$}
\label{app:polar}
To calculate the longitudinal polarization fraction $F_L$, we construct the helicity amplitude for $D^0\to V_1V_2\to(P_1P_2)(P_3P_4)$ processes as follows:

\begin{equation}
    \begin{split}
        H_{\lambda_{V_1},\lambda_{V_2}} = & \ 
        F_{\lambda_{V_1},\lambda_{V_2}}^{J_D} 
        D_{\lambda_{D},\lambda_{V_1}-\lambda_{V_2}}^{J_D}(\phi_{V_1},\theta_{V_1},0) \cdot P_{V_1} \\ 
        & \ 
        F_{\lambda_{P_1},\lambda_{P_2}}^{J_{V_1}} 
        D_{\lambda_{V_1},\lambda_{P_1}-\lambda_{P_2}}^{J_{V_1}}(\phi_{P_1},\theta_{P_1},0) \cdot P_{V_2} \\ 
        & \ 
        F_{\lambda_{P_3},\lambda_{P_4}}^{J_{V_2}} 
        D_{\lambda_{V_2},\lambda_{P_3}-\lambda_{P_4}}^{J_{V_2}}(\phi_{P_3},\theta_{P_3},0).
    \end{split}
\end{equation}

Here $\lambda$ denotes the helicity of the particle, which is rotation invariant. The superscript $J$ indicates the particle spin, and $\theta$ and $\phi$ are the helicity angles. $P_{V_{1,2}}$ denotes the propagator for vector mesons, $D$ is the Wigner D-function and $F_{\lambda_1,\lambda_2}$ is the helicity-coupling amplitude for $A \rightarrow BC$ as follow:
\begin{equation}
    F_{\lambda_1,\lambda_2}^{A\to BC}=\sum_{LS}\alpha_{LS}\sqrt{\frac{2L+1}{2J_A+1}}\langle L0;S\delta|J_A\delta\rangle\langle J_BJ_C;\lambda_1 -\lambda_2|S\delta\rangle q^LX_L.  \\
    \label{eq:helicity}
\end{equation}
Here, $L$ denotes the relative angular momentum between daughter particles $B$ and $C$, $S$ is the coupled spin. $\delta=\lambda_1-\lambda_2$ is the helicity difference, $q$ is the break-up momentum and $X_L$ denotes the Blatt-Weisskopf barrier. More detailed information can be found in Ref~\cite{BESIII:2022udq}. For this specific process, $J_D=0$, and $\lambda_{V_1}=\lambda_{V_2}$, which lead to three helicity states: $H_+$, $H_0$ and $H_-$. The total amplitude is therefore $|A|^2=|H_++H_0+H_-|^2$. The helicity amplitude coefficients for $D\to VV$ have also three components $\alpha_{0,1,2}$ corresponding to the angular momentum between vector mesons. Based on the above description, we obtain:
\begin{equation}
\hspace{-1cm}
\begin{gathered}
    H_{\pm}=(\sqrt{\frac{1}{3}}\alpha_0\mp\sqrt{\frac{1}{2}}\alpha_1M_{D}q_DX_1^D+\sqrt{\frac{1}{6}}\alpha_2q_D^2X_2^D)e^{\pm\phi_{P_1}}\frac{sin\theta_{P_1}}{\sqrt{2}}e^{\pm\phi_{P_3}}\frac{sin\theta_{P_3}}{\sqrt{2}}q_{V_1}X_1^{V_1}q_{V_2}X_1^{V_2}P_{V_1}P_{V_2},\\
    H_{0}=(-\sqrt{\frac{1}{3}}\alpha_0+\sqrt{\frac{2}{3}}\alpha_2q_D^2X_2^D)\gamma_{V_1}\gamma_{V_2}cos\theta_{P_1}cos\theta_{P_3}q_{V_1}X_1^{V_1}q_{V_2}X_1^{V_2}P_{V_1}P_{V_2}.\\
\end{gathered}
\end{equation}

With the following formulas, the coefficients of the covariant tensor amplitude $C_{0,1,2}$ are converted to the helicity amplitude coefficients by:
\begin{equation}
    \alpha_0=4\sqrt{3}c_0,
    \alpha_1=8\sqrt{2}e^{i(-\frac{\pi}{2})}c_1,
    \alpha_2=16\sqrt{\frac{2}{3}}e^{i(-\pi)}c_2,\\
\end{equation}
The helicity amplitudes $H_{\pm}$ and $H_0$ are obtained for each event by using the coefficients $c_{0,1,2}$ obtained from the amplitude analysis. The longitudinal polarization fraction $F_L$ is defined as:
\begin{equation}
\label{aaa}
    F_L=\frac{\int|H_0|^2d\Phi}{\int|H_-+H_0+H_+|^2d\Phi}=\frac{\sum_{MC}|H_0|^2}{\sum_{MC}|H_-+H_0+H_+|^2}.
\end{equation}
The integral in Eq.~(\ref{aaa}) is calculated using PHSP Monte Carlo events without considering the detection efficiency. The statistical uncertainties for $F_L$ are estimated by varying the coefficients $c_{0,1,2}$ according to the correlation matrix. A Gaussian function is then employed to fit the $F_L$ distribution, and the width is assigned as the uncertainty.

\section{ Other Intermediate States Tested}\label{app:test}
Table~\ref{tab:test} lists the amplitudes we tested but not included in the nominal fit. The significance of each
amplitude, with respect to the nominal fit are all less than 5$\sigma$.

\begin{table}[htbp]
 
  \centering
  \begin{tabular}{l c}
    \hline
    \hline
    Tested amplitude &Significance ($\sigma$)\\
    \hline
    $D^0\to K^{*}(892)^+(K^-\pi^0)_{S-\rm{wave}}$ & 1.4\\
    $D^0\to K^{*}(892)^-(K^+\pi^0)_{S-\rm{wave}}$ & 1.3\\
    $D^0\to K^{*}_0(1430)^+K^{*}_0(1430)^-$ &$<1$\\
    $D^0\to K^{*}(892)^+K^{*}_0(1430)^-$ & $<1$\\
    $D^0\to K^{*}(892)^-K^{*}_0(1430)^+$ & $<1$\\
    $D^0\to K^{*}(892)^+K^{*}(1410)^-[S]$ & 2.4\\
    $D^0\to K^{*}(892)^+K^{*}(1410)^-[P]$ & $<1$\\
    $D^0\to K^{*}(892)^+K^{*}(1410)^-[D]$ & 1.9\\
    $D^0\to K^{*}(892)^-K^{*}(1410)^+[S]$ & $<1$\\
    $D^0\to K^{*}(892)^-K^{*}(1410)^+[P]$ & 1.8\\
    $D^0\to K^{*}(892)^-K^{*}(1410)^+[D]$ & 2.9\\
    $D^0\to K_1(1400)^-K^+, K_1(1400)^-\to K^{*}(892)^-\pi^0$ &2.9\\
    $D^0\to K_1(1270)^+K^-, K_1(1270)^+\to K^{*}(892)^+\pi^0$ &2.5\\
    $D^0\to K_1(1270)^-K^+, K_1(1270)^-\to K^{*}(892)^-\pi^0$ &2.7\\
    $D^0\to K^*(1410)^-K^+, K^*(1410)^-\to K^{*}(892)^-\pi^0$ &1.3\\
    $D^0\to K^*(1410)^+K^-, K^*(1410)^+\to K^{*}(892)^+\pi^0$ &2.7\\
    $D^0\to K_1(1270)^-K^+, K_1(1270)^-\to (K^-\pi^0)_{S-\rm{wave}}\pi^0$ &1.7\\  
    $D^0\to K_1(1400)^+K^-, K_1(1400)^+\to (K^+\pi^0)_{S-\rm{wave}}\pi^0$ &1.2\\
    $D^0\to K_1(1270)^-K^+, K_1(1270)^-\to (K^-\pi^0)_{S-\rm{wave}}\pi^0$ &1.6\\
    $D^0\to K_1(1270)^+K^-, K_1(1270)^+\to (K^+\pi^0)_{S-\rm{wave}}\pi^0$ &1.3\\
    $D^0\to K^*(1410)^-K^+, K^*(1410)^-\to (K^-\pi^0)_{S-\rm{wave}}\pi^0$ &$<1$\\
    $D^0\to K^*(1410)^+K^-, K^*(1410)^+\to (K^+\pi^0)_{S-\rm{wave}}\pi^0$ &$<1$\\
    $D^0\to f_1(1420)\pi^0, f_1(1420)[D]\to K^{*}(892)^+K^-$ &$<1$\\
    $D^0\to f_1(1420)\pi^0, f_1(1420)[D]\to K^{*}(892)^-K^+$ &$<1$\\
    $D^0\to f_1(1420)\pi^0, f_1(1420)\to (K^+\pi^0)_{S-\rm{wave}}K^-$ &2.2\\
    $D^0\to f_1(1420)\pi^0, f_1(1420)\to (K^-\pi^0)_{S-\rm{wave}}K^+$& 1.0\\
    $D^0\to f_1(1510)\pi^0, f_1(1510)[S]\to K^{*}(892)^+K^-$ &1.4\\
    $D^0\to f_1(1510)\pi^0, f_1(1510)[S]\to K^{*}(892)^-K^+$ &1.3\\
    $D^0\to f_1(1510)\pi^0, f_1(1510)[D]\to K^{*}(892)^+K^-$ &1.1\\
    $D^0\to f_1(1510)\pi^0, f_1(1510)[D]\to K^{*}(892)^-K^+$ &$<1$\\
    $D^0\to f_1(1510)\pi^0, f_1(1510)\to (K^+\pi^0)_{S-\rm{wave}}K^-$ &1.8\\
    $D^0\to f_1(1510)\pi^0, f_1(1510)\to (K^-\pi^0)_{S-\rm{wave}}K^+$& 1.0\\
    $D^0\to \eta(1295)\pi^0, \eta(1295)\to K^{*}(892)K$ &$<1$\\
    $D^0\to \eta(1475)\pi^0, \eta(1475)\to K^{*}(892)K$ &$<1$\\
    $D^0\to \eta(1295)\pi^0, \eta(1295)\to a_0(980)\pi^0$ &2.8\\  
    $D^0\to \eta(1475)\pi^0, \eta(1475)\to a_0(980)\pi^0$ &2.2\\
    $D^0\to \eta(1475)\pi^0, \eta(1475)\to (K^+\pi^0)_{S-\rm{wave}}K^-$ &2.2\\
    $D^0\to \eta(1475)\pi^0, \eta(1475)\to (K^-\pi^0)_{S-\rm{wave}}K^+$ &2.7\\   
    $D^0\to \phi(1680)\pi^0, \phi(1680)\to K^{*}(892)^+K^-$ &$<1$\\
    $D^0\to \phi(1680)\pi^0, \phi(1680)\to K^{*}(892)^-K^+$ &$<1$\\
    \hline
    \end{tabular}
    \caption{Tested amplitudes, but not included in the nominal fit.\label{tab:test}}
\end{table}

\section{Interference fit fractions}
\label{app:int}
Table~\ref{tab:intfitfrac} provides the interference fit fractions for each pair of contributions to the amplitude model.

\begin{landscape}

\begin{table}

    \centering
        \begin{tabular}{|c|cccccccc|}
        \hline
           &II&III&IV&V&VI&VII&VIII&IX  \\
                       
        \hline
         I     &0.05$\pm$0.01  &0.93$\pm$0.28  &0.05$\pm$0.01  &0.06$\pm$0.02 &$-$0.76$\pm$0.95  &0.97$\pm$0.56 &0.91$\pm$0.51  &2.93$\pm$0.34\\
         II    &&0.00$\pm$0.00  &0.00$\pm$0.00  &0.00$\pm$0.00 &0.01$\pm$0.00  &$-$0.01$\pm$0.01 &0.00$\pm$0.01  &0.00$\pm$0.01 \\
         
         III   &&&$-$0.47$\pm$0.18&$-$0.47$\pm$0.18&0.27$\pm$0.22&1.30$\pm$0.47&1.30$\pm$0.46&$-$0.84$\pm$0.18 \\
         IV    &&&& 1.91$\pm$0.48 & $-$0.12$\pm$0.09 &0.47$\pm$0.07 &$-$0.09$\pm$0.03&1.09$\pm$0.42 \\
         V     &&&&&  $-$0.15$\pm$0.08& $-$0.09$\pm$0.02 &0.48$\pm$0.07&0.22$\pm$0.09\\
         VI    &&&&&&  $-$2.50$\pm$0.34&$-$2.50$\pm$0.34 &$-$7.39$\pm$1.18\\
         VII   &&&&&&&  2.73$\pm$0.38&$-$1.01$\pm$0.20  \\
         VIII  & & & & & & & &   $-$1.05$\pm$0.21 \\
        \hline
    
        \end{tabular}
        
       \caption{ Interference of each amplitude, in unit of \% of total amplitude. I $D^0[S]\to K^{*}(892)^+ K^{*}(892)^-$, II $D^0[P]\to K^{*}(892)^+ K^{*}(892)^-$,III $D^0[D]\to K^{*}(892)^+ K^{*}(892)^-$, IV $D^0\to f_1(1420)\pi^{0}, f_1(1420)\to K^{*}(892)^+ K^{-}$, V $D^0\to f_1(1420)\pi^{0}, f_1(1420)\to K^{*}(892)^- K^{+}$. VI $D^0\to \eta(1405)\pi^{0}, \eta \to a_{0}(980) \pi^{0}$, VII $D^0\to \eta(1475)\pi^{0}, \eta(1475)\to K^{*}(892)^+ K^{-}$, VIII $D^0\to \eta(1475)\pi^{0}, \eta(1475)\to K^{*}(892)^- K^{+}$, IX $D^0\to K_1(1400)^+ K^{-}$.  \label{tab:intfitfrac}}    

    \end{table}
\end{landscape}

\clearpage

M.~Ablikim$^{1}$\BESIIIorcid{0000-0002-3935-619X},
M.~N.~Achasov$^{4,d}$\BESIIIorcid{0000-0002-9400-8622},
P.~Adlarson$^{81}$\BESIIIorcid{0000-0001-6280-3851},
X.~C.~Ai$^{87}$\BESIIIorcid{0000-0003-3856-2415},
C.~S.~Akondi$^{31A,31B}$\BESIIIorcid{0000-0001-6303-5217},
R.~Aliberti$^{39}$\BESIIIorcid{0000-0003-3500-4012},
A.~Amoroso$^{80A,80C}$\BESIIIorcid{0000-0002-3095-8610},
Q.~An$^{77,64,\dagger}$,
Y.~H.~An$^{87}$\BESIIIorcid{0009-0008-3419-0849},
Y.~Bai$^{62}$\BESIIIorcid{0000-0001-6593-5665}
% ...
\\
\vspace{0.2cm}
(BESIII Collaboration)\\
\vspace{0.2cm} {\it
$^{1}$ Institute of High Energy Physics, Beijing 100049, People's Republic of China\\
% ...
$^{4}$ Budker Institute of Nuclear Physics SB RAS (BINP), Novosibirsk 630090, Russia\\
% ...
$^{31}$ INFN Sezione di Ferrara, (A)INFN Sezione di Ferrara, I-44122, Ferrara, Italy; (B)University of Ferrara, I-44122, Ferrara, Italy\\
% ...
$^{39}$ Johannes Gutenberg University of Mainz, Johann-Joachim-Becher-Weg 45, D-55099 Mainz, Germany\\
% ...
$^{62}$ Southeast University, Nanjing 211100, People's Republic of China\\
% ...
$^{64}$ State Key Laboratory of Particle Detection and Electronics, Beijing 100049, Hefei 230026, People's Republic of China\\
% ...
$^{77}$ University of Science and Technology of China, Hefei 230026, People's Republic of China\\
% ...
$^{80}$ University of Turin and INFN, (A)University of Turin, I-10125, Turin, Italy; (B)University of Eastern Piedmont, I-15121, Alessandria, Italy; (C)INFN, I-10125, Turin, Italy\\
$^{81}$ Uppsala University, Box 516, SE-75120 Uppsala, Sweden\\
% ...
$^{87}$ Zhengzhou University, Zhengzhou 450001, People's Republic of China\\

\vspace{0.2cm}
$^{\dagger}$ Deceased\\
% ...
$^{d}$ Also at the Novosibirsk State University, Novosibirsk, 630090, Russia\\
% ...

}

\end{document}